\documentclass[a4paper,aps,pre,floatfix,twocolumn, superscriptaddress,nobibnotes]{revtex4-1}

\usepackage[pdftex]{graphicx}
\setkeys{Gin}{width=0.9\columnwidth}

\usepackage{times}
\usepackage{verbatim}
\usepackage{bbold}
\usepackage{bbm}

\usepackage{amsmath}
\usepackage{amsfonts}
\usepackage{color}

\usepackage[english]{babel}
\usepackage{microtype}
\usepackage{comment}
\usepackage[normalem]{ulem}
\usepackage{float}

\usepackage{relsize}

\usepackage[hidelinks,unicode=true]{hyperref}
\hypersetup{colorlinks=true,linkcolor=blue,urlcolor=blue,citecolor=blue,pdfhighlight=/N
}

\usepackage{lipsum}

\usepackage[title]{appendix}

\usepackage{latexsym}
\usepackage{xcolor}
\usepackage{fancyhdr}
\usepackage{tabularx}
\usepackage{gensymb}
\newcolumntype{Y}{>{\centering\arraybackslash}X}

\newcommand{\av}[1]{\langle #1 \rangle}

\renewcommand{\sout}[1]{\unskip}

\newcommand{\add}[1]{{#1}}
\newcommand{\addd}[1]{{#1}}
\newcommand{\delete}[1]{\sout{#1}}
\newcommand{\addnew}[1]{{{#1}}}
\newcommand{\delnew}[1]{\sout{#1}}

\begin{document}


\title{Modeling Explosive Opinion Depolarization in Interdependent Topics}

\author{Jaume Ojer}
\affiliation{Departament de F\'{\i}sica, Universitat Polit\`ecnica de Catalunya, Campus Nord B4, 08034 Barcelona, Spain}

\author{Michele Starnini}
\email{Corresponding author: michele.starnini@gmail.com}
\affiliation{Departament de F\'{\i}sica, Universitat Polit\`ecnica de Catalunya, Campus Nord B4, 08034 Barcelona, Spain}
\affiliation{CENTAI, Turin, Italy}

\author{Romualdo Pastor-Satorras}
\email{Corresponding author: romualdo.pastor@upc.edu}
\affiliation{Departament de F\'{\i}sica, Universitat Polit\`ecnica de Catalunya, Campus Nord B4, 08034 Barcelona, Spain}

\date{\today}

\begin{abstract}
Understanding the dynamics of opinion depolarization is pivotal to reducing the political divide in our society. We propose an opinion dynamics model, \addd{which we name the social compass model,} for interdependent topics represented in a polar space, where zealots holding extreme opinions are less prone to change their minds. We analytically show that the phase transition from \delete{initial} polarization to \delete{a} consensus, as a function of increasing social influence, is explosive if topics are not correlated. We validate our theoretical framework \delete{with empirical opinion polls} \add{through extensive numerical simulations and recover explosive depolarization also by using initial opinions} from the American National Election Studies, including polarized and interdependent topics.
\end{abstract}

\maketitle



The presence of opinion polarization---i.e., two groups holding opposite and possibly extreme opinions in a population---has been extensively observed with respect to several controversial topics, ranging from religion~\cite{perry_american_2022} and race~\cite{montalvo_ethnic_2005} to climate change~\cite{mccright2011politicization} and political ideology~\cite{mccoy_polarization_2018}. Polarization may contribute to deepening the political divide in our society~\citep{iyengar2012affect}, hampering the collective resolution of important societal challenges~\cite{wang2020communicating}, and even fostering the spreading of misinformation and conspiracy theories~\cite{michael2017rise}. 
Consequently, an interest towards a theoretical understanding of the emergence of opinion polarization has arisen in several disciplines, from statistical physics to social and computer science.

Models that reproduce polarization are based on different opinion dynamics mechanisms, such as homophily~\cite{dandekar_biased_2013, baumann19}, bounded confidence~\cite{deffuant_mixing_2000, hegselmann_opinion_2002, lorenz_continuous_2007}, or opinion rejection~\cite{huet_rejection_2008, crawford_opposites_2013}. Modeling the process of reducing opinion polarization amongst the population, or \emph{depolarization}~\cite{vinokur1978depolarization}, has also been the object of recent work~\cite{matakos_measuring_2017, musco_minimizing_2018, balietti_reducing_2021}.
In most cases, such modeling efforts address the simplest case of one-dimensional opinions with respect to a single topic~\cite{jager_uniformity_2005, chau_social_2014}.
However, the process of opinion formation may invest multiple topics at the same time~\cite{poole_spatial_2005, benoit_dimensionality_2012}, requiring a proper multi-dimensional modeling framework for opinion dynamics~\cite{li_agent-based_2017, van_der_maas_polarization_2020, schweighofer_agent-based_2020, chen_modeling_2021}. 
When multiple topics are taken into account, a crucial feature can often be observed: issue alignment~\cite{dimaggio_have_1996, baldassarri_partisans_2008}, i.e., the presence of correlations between opinions with respect to different topics, especially along the so-called left-right dimension~\cite{freire_party_2008, falck_measuring_2020}.
For instance, individuals with strong religious beliefs are more likely to oppose abortion legalization~\cite{adamczyk_religion_2022}, while other non-trivial correlations can emerge~\cite{baldassarri_partisans_2008, benoit_dimensionality_2012, dellaposta_why_2015}.
\delete{Strikingly, most} \add{However, many} multi-dimensional opinion models proposed so far failed to reproduce opinion polarization~\cite{laguna_vector_2003, fortunato_vector_2005, etesami_termination_2013}, neglecting the interdependence among different topics~\cite{converse_nature_2006, baumann_emergence_2021}.

In this Letter, we aim to fill this gap 
by proposing an analytically tractable model of opinion dynamics in a space of two interdependent topics.
We represent this space in the polar plane, where the angle represents the orientation of an individual with respect to \delete{two} \add{both} topics, and the radius expresses the attitude strength (referred to as conviction in the literature~\cite{abelson_conviction_1988}). 
This polar representation allows naturally to formulate the key assumption of the model, i.e., zealots with extreme opinions (large conviction) may be less prone to change their opinion than individuals with small conviction, in line with experimental psychology~\cite{miller_attitude_1993, pomerantz_attitude_1995}.
We observe that this model, \addd{which we name the \emph{social compass model}}, exhibits a phase transition
from an initial polarized state to a depolarized or consensus one, as a function of increasing social influence.
We analytically show \addnew{at the mean-field level} that the nature of such transition depends on the correlation between initial opinions: uncorrelated opinions trigger a first-order, or explosive, depolarization to consensus, while correlated initial opinions lead to a second-order, continuous transition.
We \delete{validate} \add{test} our theoretical framework \delete{with} \add{by using} real data of polarized \add{initial} opinions with respect to interdependent topics.

We start by defining a representation of opinions in polar space.
Let us consider $N$ individuals, each agent $i$ holding opinions $(x_i, y_i)$ towards two distinct topics $X$ and $Y$, respectively, that are assumed to be normalized in the interval $x_i$,$y_i$ $\in [-1,1]$. 
The combined opinion of each individual with respect to the two topics can be represented in polar coordinates by 
its conviction $\rho_i = \sqrt{x_i^2 + y_i^2}$ and its orientation $\varphi_i = \arctan{(y_i/x_i)}$, with $\varphi_i \in [-\pi, \pi]$.
For instance, two agents $i$ and $j$ holding extreme and opposite opinions, $x_i = y_i = 1$, $x_j = y_j = -1$, will be represented in the polar plane with the same, maximum conviction $\rho_i = \rho_j = \sqrt{2}$ and opposite orientations $\varphi_i = \pi/4$ and $\varphi_j = -3\pi/4$.
Note that representing individuals in a plane defined by two major axes, such as libertarian vs. authoritarian within a social context, and left vs. right within an economic context, is not novel in political science~\footnote{\url{http://www.politicalcompass.org}}.

To support this polar representation, we consider empirical opinions from the American National Election Studies (ANES)~\footnote{\url{http://www.electionstudies.org}}, see Supplementary Material Section~I (SM~I). 
The angular distribution $P(\varphi)$ obtained from the ANES dataset for different pairs of topics shows a rich phenomenology. 
We highlight four interesting cases, reported in Supplementary Figure SF~1 of SM~IA. 
First, there can be no consensus with respect to either topic, with opinions roughly uniformly distributed, so that $P(\varphi)$ will also be a uniform distribution, see SF~1(a). 
Second, a consensus around both topics may emerge, indicated by the $P(\varphi)$ distribution peaked around a certain consensus value $\varphi^\ast = \arctan(y^\ast/x^\ast)$, where $y^\ast$ and $x^\ast$ are the consensus opinions of topics $Y$ and $X$, respectively, see SF~1(b). Third, opinions with respect to both topics can be polarized, i.e., both one-dimensional opinion distributions are characterized by two well-separated peaks. Here we can distinguish two different cases: opinions can be polarized but not correlated, 
for which the $P(\varphi)$ distribution will be characterized by four peaks (quadrimodal distribution), representing for example the four extreme combinations $(x_i,y_i) = (+1,+1),(+1,-1),(-1,+1),(-1,-1)$, see SF~1(c).
Finally, opinions can be polarized \emph{and} strongly correlated, 
shown in SF~1(d). In this case, one can only observe two peaks in the $P(\varphi)$ (bimodal distribution), corresponding to two ideological combinations like $(x_i,y_i) = (+1,+1),(-1,-1)$. 


Within this context, we study how social influence can affect the initial opinions of individuals 
by proposing \addd{the social compass model}, which is inspired by the Friedkin-Johnsen model~\cite{friedkin_social_1990}.
For each individual $i$, we focus on the time evolution of their orientation, represented by $\theta_i(t)$, provided their initial orientation $\theta_i(0) = \varphi_i$ and that their conviction $\rho_i$ will not change over time.
We rely on only two key assumptions: 
i) agents exert a certain degree of social influence on their peers, and 
ii) each agent $i$ has a tendency to maintain their initial opinion $\varphi_i$ proportional to their conviction $\rho_i$ (i.e., agents with high conviction are more stubborn).
We operationalize this simple theoretical framework in the following set of $N$ ordinary differential equations,
\begin{equation}
    \dot{\theta}_i(t) = \rho_i \sin \left[ \varphi_i - \theta_i(t) \right] + \frac{\lambda}{N} \sum_{j = 1}^N \sin \left[ \theta_j(t) - \theta_i(t) \right],
    \label{eq:model}
\end{equation}
where $\lambda$ is a coupling constant that quantifies the strength of social influence. 
\addd{In a real-world scenario, social influence, indicating the tendency of individuals to adjust their behavior to meet the expectations of their peers, could be empirically quantified by surveys.}
\delnew{and where} We assume that each individual can interact with all other individuals, \addnew{which allows us to solve the model through a mean-field approach.}
Since opinions are described by angles, it is natural to model consensus formation as the alignment of the agents' orientations~\cite{hegselmann_opinion_2002, pluchino_changing_2005, castellano_statistical_2009}, \add{with a phase coupling inspired in the Kuramoto model~\cite{kuramoto_self-entrainment_1975}.}

\addd{The social compass model} exhibits a phase transition at a threshold value $\lambda_c$ of the coupling constant, separating a polarized from a depolarized (consensus) phase. 
This transition can be characterized in terms of the order parameter $r$, defined by~\cite{Acebron.2005}
\begin{equation}
    r(\lambda)\mathrm{e}^{i\psi(\lambda)} = \frac{1}{N} \sum_{j = 1}^N \mathrm{e}^{i\theta_j(\lambda)},
    \label{eq:order_parameter}
\end{equation}
where $\theta_j(\lambda)$ is computed at the steady state of Eq.~\eqref{eq:model} and $\psi(\lambda)$ is the average orientation in the population. 
In the absence of social interactions ($\lambda = 0)$, Eq.~\eqref{eq:model} leads to the steady state $\theta_j = \varphi_j$. 
In accordance with the empirical evidence presented above, we are interested in initial states with polarized orientations following bimodal or quadrimodal $P(\varphi)$ distributions. For this polarized state, we have $r(\lambda= 0) = 0$ provided that $\av{\cos{(\varphi)}} = \av{\sin{(\varphi)}} = 0$, where $\av{...}$ denotes the average value. 
For sufficiently large $\lambda$, a consensus state leads to $r \simeq 1$, meaning that all the agents have the same average orientation $\theta_j \simeq \psi$.

Given that the initial opinions are distributed according to uncorrelated probability densities $P(\rho)$ and $P(\varphi)$, a general solution of the model can be found by the self-consistent equation $r = I(r, \psi)$ (see SM~II)~\cite{Acebron.2005}, with
\begin{equation}
    I(r, \psi) = \int_0^\infty \mkern-15mu d\rho \int_{-\pi}^{\pi} \mkern-15mu d\varphi \frac{P(\rho) P(\varphi) \left[ \lambda r + \rho \, \mathrm{e}^{i \left( \varphi - \psi \right)} \right]}{\sqrt{(\lambda r)^2 + 2\lambda r \rho \cos(\varphi - \psi ) + \rho^2}},
    \label{eq:general}
\end{equation}
which translates into the pair of equations $r = \mathrm{Re} \{ I(r, \psi) \}$ and $0 = \mathrm{Im} \{ I(r, \psi) \}$ for the real and imaginary parts of $I(r, \psi)$, respectively. The equation for the imaginary part is used to identify the average orientation $\psi$, which, plugged into the equation for the real part, allows to compute the order parameter $r$ as a function of $\lambda$.
We can establish a threshold condition for the depolarized state considering the instability of the solution $r=0$, which translates into the condition $\partial \mathrm{Re} \{ I(r, \psi) \} / \partial r \vert_{r=0} \geq 1$. This leads to a depolarized state for $\lambda > \lambda_c$ with (see SM~IIA)
\begin{equation}
    \lambda_c = \frac{1}{\int_0^\infty d\rho \frac{P(\rho)}{\rho} \int_{-\pi}^{\pi}d\varphi P(\varphi) \sin^2(\varphi - \psi)}.
    \label{eq:threshold}
\end{equation}

We \delete{have} tested the validity of this theoretical result by using empirical data from the ANES opinion polls as values of the initial orientation and conviction, $P(\varphi)$ and $P(\rho)$, respectively. 
We \delete{have} focused in particular on correlated polarized opinions represented by bimodal $P(\varphi)$, reported in SF~3. As we can see in Fig.~\ref{fig:empirical_threshold}, in all cases we observe numerically a depolarization transition at a threshold that is approximately described by the theoretical prediction Eq.~\eqref{eq:threshold}. 

\begin{figure}[t]
    \centering
    \includegraphics[width=0.9\columnwidth]{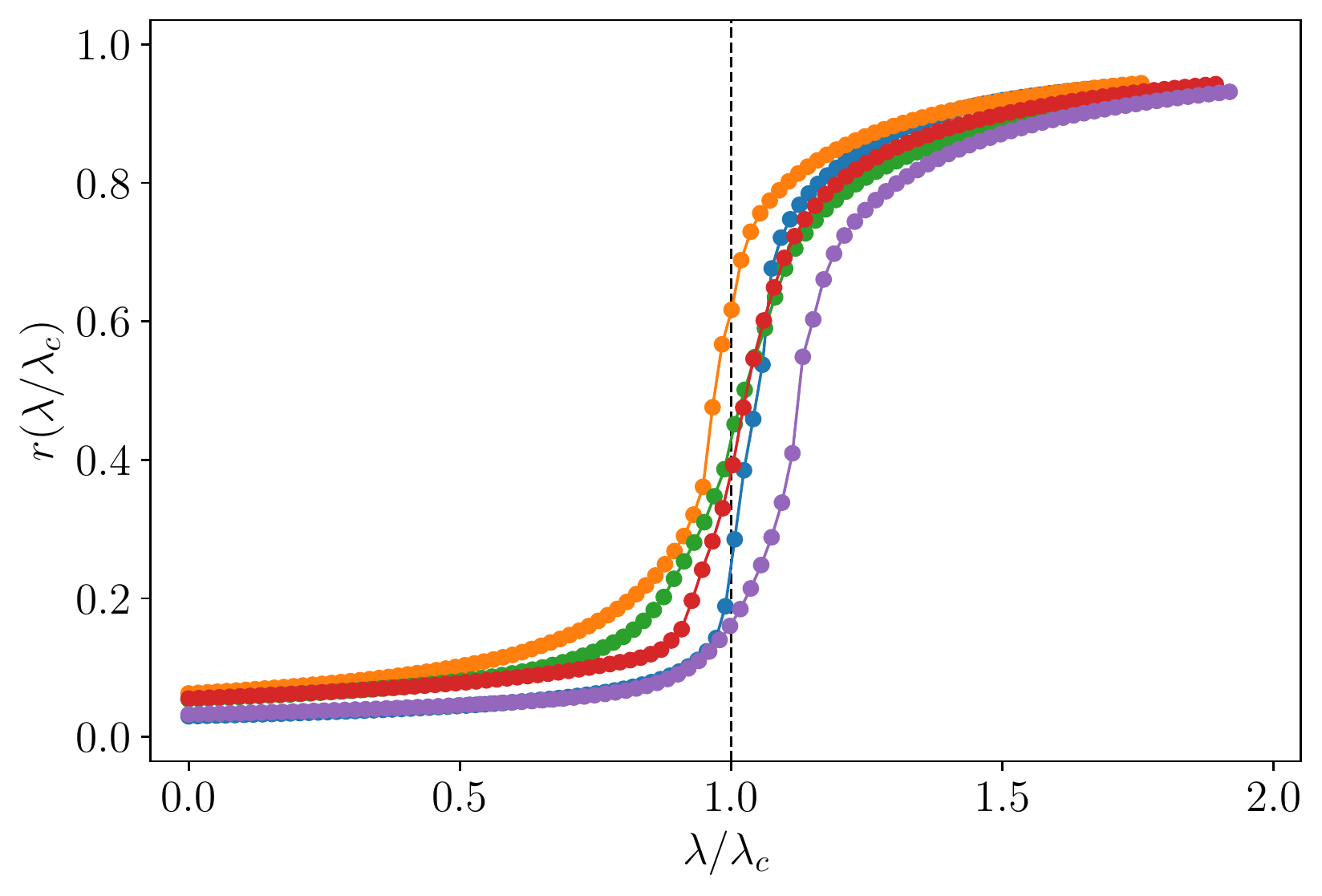}
    \caption{Order parameter $r(\lambda/\lambda_c)$, by using different initial opinion distributions from ANES data (topics described in the SF~3). The theoretical threshold $\lambda_c$ is computed numerically from Eq.~\eqref{eq:threshold}, by using the empirical distributions $P(\rho)$ and $P(\varphi)$, 
    see SM~IIB.
    }
    \label{fig:empirical_threshold}
\end{figure}

To address analytically the nature of the observed depolarization transition, we consider specific forms of the conviction and orientation distributions. From the ANES dataset, the conviction distribution shows in general an increasing trend, see SM~IB.
 We thus model it as a power-law form,
\begin{equation}
    P(\rho) = (\alpha + 1) \rho^\alpha,
    \label{eq:conviction}
\end{equation}
where $\alpha = 0$ corresponds to a uniform distribution, and the limit  $\alpha \to \infty$ represents the case constant conviction, $P(\rho) = \delta(\rho - \rho_\mathrm{max})$. We arbitrarily fix $\rho_\mathrm{max} = 1$.
Regarding the orientation distribution, we choose the general form
\begin{eqnarray}
    P(\varphi) &=& \frac{1}{4} \left[ \delta(\varphi - \varphi_0) + \delta(\varphi - \varphi_0 + \pi) \right] \nonumber \\
    &+& \frac{1}{4} \left[ \delta(\varphi + \varphi_0) + \delta(\varphi + \varphi_0 - \pi) \right], 
    \label{eq:angular}
\end{eqnarray}
fulfilling the initial polarized state condition $\av{\cos(\varphi)} = \av{\sin(\varphi)} = 0$.
 For symmetry reasons, we restrict $\varphi_0 \in [0, \pi/4]$ and consider separately the cases of correlated (bimodal distribution, $\varphi_0 = 0$) and uncorrelated (quadrimodal distribution, $\varphi_0 >0$)  polarization.
The analysis of the equation $0 = \mathrm{Im} \{ I(r, \psi) \}$ leads in both cases to an average orientation $\psi = \pm \pi/2$ independent of $\lambda$ (see SM~III and SM~IV), which we will impose in the following analysis.
\addd{We note that the average orientation $\psi$ in the depolarized
phase falls exactly at the middle point between the two peaks.
Therefore, the consensus emerges as a positive solution where no initial opinion dominates over the other.}

\begin{figure}[t]
    \centering
    \includegraphics[width=0.9\columnwidth]{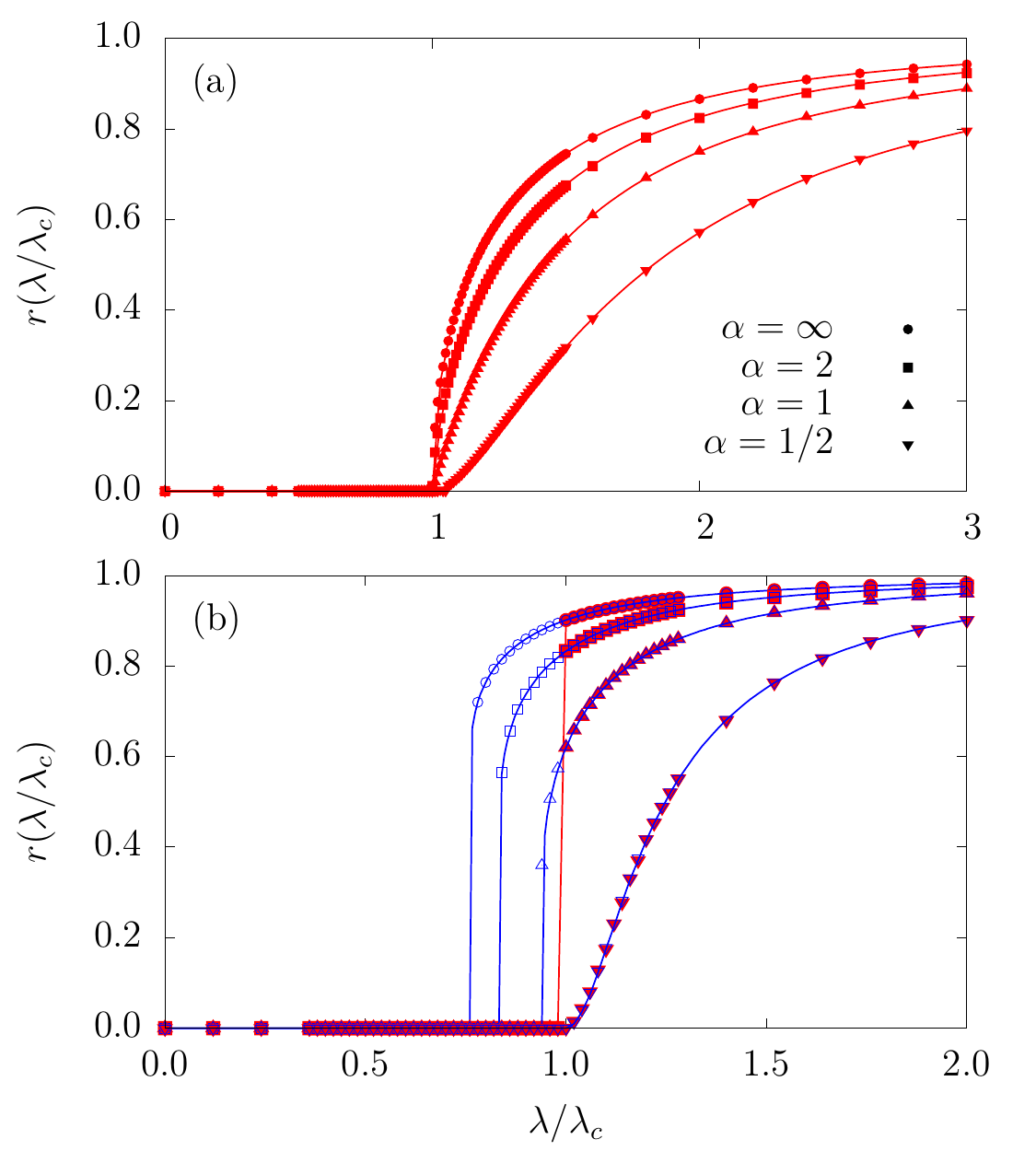}
    \caption{Order parameter $r(\lambda / \lambda_c)$ for different values of $\alpha$. 
    The initial orientation distribution $P(\varphi)$ is polarized and correlated (panel (a), $\varphi_0 = 0$) or fully uncorrelated (panel (b), $\varphi_0 = \pi/4$). 
    Points (solid lines) represent numerical simulations (theoretical predictions). 
    System size $N = 10^6$. 
    For the uncorrelated case, backward continuation (blue) is plotted in empty symbols. 
    Curves of forward continuation (red) for different $\alpha$ collapse for $\lambda \leq \lambda_c$, while they are identical to backward continuations for $\lambda > \lambda_c$.
    }
    \label{fig:rhoalpha}
\end{figure}

In the case of correlated and polarized initial opinions, corresponding to a bimodal $P(\varphi)$, the function $\mathrm{Re} \{ I(r, \psi) \}$ can be integrated analytically, yielding the self-consistent equation
\begin{equation}
    r =  \, _2F_1 \left( \frac{1}{2}, \frac{\alpha + 1}{2}; \frac{\alpha + 3}{2}; -\frac{1}{(\lambda r)^2} \right),
    \label{eq:self_general}
\end{equation}
where $_2F_1(a,b;c;z)$ is the Gaussian hypergeometric function~\cite{abramovitz}. To study the behavior in the vicinity of the depolarization transition, we perform a leading order expansion of the hypergeometric function around $r=0$ and solve the ensuing equation (see SM~IIIA). 
We obtain that the transition is continuous, with the typical behavior of the order parameter $r(\lambda) \sim (\lambda -\lambda_c)^\beta$, with a threshold and an exponent $\beta$ depending on $\alpha$ as
\begin{equation}
    \lambda_c(\alpha) = \frac{\alpha}{\alpha + 1}, \quad \beta(\alpha) = 
    \begin{cases}
        1/\alpha & \text{if } \alpha < 2 \\
        1/2 & \text{if } \alpha > 2
    \end{cases}.
    \label{eq:threshold_alpha}
\end{equation}
The particular values $\alpha=1$ and $\alpha \to \infty$ can be solved analytically, recovering the asymptotic result of Eq.~\eqref{eq:threshold_alpha} (see SM~IIIA1).
Fig.~\ref{fig:rhoalpha}(a) shows numerical simulations of $r$ compared with the theoretical prediction from the numerical or analytical solution of Eq.~\eqref{eq:self_general}. 
The match obtained is perfect, confirming a continuous transition in this case. 
We estimated the values of the exponent $\beta$ by performing a linear regression of $r$ as a function of $\lambda - \lambda_c$ in the vicinity of the transition. 
The values obtained, reported in Table~\ref{tab:exponents}, confirm the validity of our theoretical approach.

\begin{table}[t]
    \begin{ruledtabular}
        \begin{tabular}{ccccccc}
            $\alpha$ & $1/3$ & $1/2$ & $1$ & $2$ & $3$ & $\infty$ \\
            \hline
            $\beta$ & $2.74(1)$ & $1.96(1)$ & $0.99(1)$ & $0.57(1)$ & $0.51(1)$ & $0.50(1)$ \\
        \end{tabular}
    \end{ruledtabular}
    \caption{Numerical exponent $\beta$ for a system size $N = 10^6$ and different values of $\alpha$, estimated from a linear regression of a double logarithmic plot of $r$ as a function of $\lambda - \lambda_c$  in the vicinity of the threshold.
    Numerical estimation is in good agreement with the theoretical prediction given by Eq.~\eqref{eq:threshold_alpha}.
    Deviations from it (as for $\alpha=1/3$) could be ascribed to finite size effects.
    }
    \label{tab:exponents}
\end{table}

For \delete{initial} polarized, but not correlated, \add{initial} opinions, $P(\varphi)$ corresponds to a quadrimodal distribution given by Eq.~\eqref{eq:angular} with $\varphi_0 > 0$. The resulting integral of $\mathrm{Re} \{ I(r, \psi) \}$ does not allow an analytical treatment, so we resort to solving numerically the corresponding self-consistent equation, see SM~IVA.  
Fig.~\ref{fig:rhoalpha}(b) shows numerical simulations of $r$ (symbols) compared with the corresponding numerical solution (lines) of the self-consistent equation for the quadrimodal symmetric case with $\varphi_0 = \pi/4$. 
For small values of $\alpha$ we observe a continuous transition, as in the bimodal case. 
However, for sufficiently large $\alpha$, the transition becomes discontinuous, i.e., we observe an explosive depolarization. The first-order nature of the transition is reflected in the presence of hysteresis observed when performing forward and backward continuation experiments, see SM~V. 

\begin{figure}[t]
    \centering
    \includegraphics[width=0.9\columnwidth]{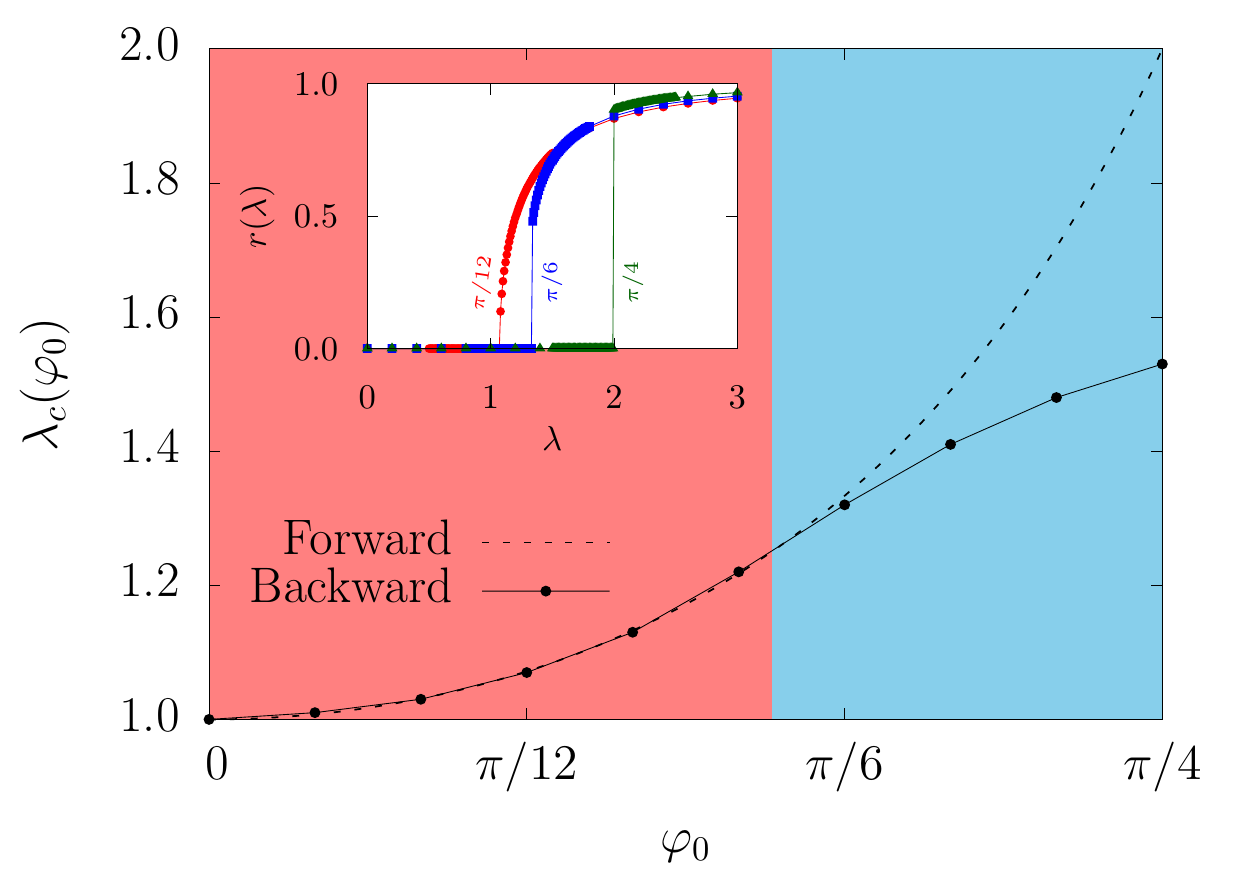}
    \caption{Inset: Order parameter $r(\lambda)$ in a forward continuation for different values of $\varphi_0$. 
    Points (solid lines) represent numerical simulations (theoretical predictions).
    Main: Threshold value $\lambda_c$ as a function of $\varphi_0$.
    We compare it by following a forward (dashed line, $\lambda_c$) and backward (solid line, $\lambda_u$) continuations in $\lambda$ of the phase transition, obtained as the theoretical prediction given by Eq.~\eqref{eq:threshold_angle} and numerical simulations, respectively.
    The plot is colored in red (blue) for $\varphi_0 < \varphi_c$ ($\varphi_0 > \varphi_c$), signaling the theoretical separation between continuous and explosive transitions. 
    System size $N = 10^3$, constant conviction $\rho = 1$ ($\alpha = \infty$).}
    \label{fig:angular}
\end{figure}

We further explore the nature of this discontinuous transition by focusing on the case in which all agents hold maximum conviction ($\alpha = \infty$).
From Eq.~\eqref{eq:threshold}, we can compute analytically the transition threshold $\lambda_c$ depending on the symmetry of the initial orientation distribution $P(\varphi)$, namely the angle $\varphi_0$,
\begin{equation}
    \lambda_c(\varphi_0) = \frac{1}{\cos^2{(\varphi_0)}}.
    \label{eq:threshold_angle}
\end{equation}
The nature of the transition can be uncovered by performing a Taylor expansion on the right-hand-side of the self-consistent equation $r = \mathrm{Re} \{ I(r, \psi) \}$ for $r$ small in the vicinity of $\lambda_c(\varphi_0)$, and solving for the analytic continuation of the solution $r = 0$. A non-zero solution for $\lambda > \lambda_c$ is indicative of a continuous transition, while for $\lambda < \lambda_c$ corresponds to the unstable branch of a discontinuous transition. This analysis shows the presence of a threshold angle $\varphi_c = \arcsin{\left( 1/\sqrt{5} \right)}$, such that for $\varphi_0 < \varphi_c$ the transition is continuous, whereas it is explosive for $\varphi_0 > \varphi_c$ (see SM~IVB for details). 

In Fig.~\ref{fig:angular} (inset) we show the perfect match between the numerical simulations of $r$ (symbols) and the numerical solution of $r = \mathrm{Re} \{ I(r, \psi) \}$ (lines) for different values of $\varphi_0$ of a quadrimodal $P(\varphi)$.
One can see that the nature of the transition changes from continuous, for small values of $\varphi_0$, to discontinuous, or explosive, for large values of $\varphi_0$.
In Fig.~\ref{fig:angular} (main) we further check this change, by plotting the instability threshold $\lambda_u$ of the upper branch of the solution, as obtained by a backward continuation simulation, see SM~V.
As we can see in Fig.~\ref{fig:angular} (main), for $\varphi_0 < \varphi_c$ the threshold of the upper branch $\lambda_u$ coincides with the threshold $\lambda_c$ for the instability of the zero solution obtained by a forward continuation, indicative of a continuous transition. For $\varphi_0 > \varphi_c$, on the other hand, $\lambda_u < \lambda_c$, signaling the hysteresis typical of explosive, discontinuous transitions.

Interestingly, \delete{a clear difference between continuous and explosive transitions depending on $P(\varphi)$ is also visible {in} empirical ANES data.} \add{the theoretical predictions of the model are also recovered by starting with an initial opinion distribution extracted from empirical ANES data.}
If \delete{the initial opinion distribution} \add{it} is polarized \emph{and} correlated (approximately bimodal $P(\varphi)$), 
the depolarization is continuous.
For instance, Fig.~\ref{fig:empirical_ordertransitions}(a) shows the topics ``religion providing guidance in day-to-day living" and ``business owners are allowed to refuse services to same-sex couples if they violate their religious beliefs".
If instead topics are polarized but uncorrelated, represented by a quadrimodal $P(\varphi)$, we observe an explosive depolarization with hysteresis.
Fig.~\ref{fig:empirical_ordertransitions}(b) considers topics ``children of unauthorized immigrants born in the U.S. should automatically get citizenship" and ``the U.S. should send troops to fight Islamic militants".
These results, in full agreement with the theoretical analysis, are confirmed by other examples of polarized initial opinion distributions from the ANES dataset,
detailed in SM~VI.

\begin{figure}[t]
    \centering
    \includegraphics[width=0.9\columnwidth]{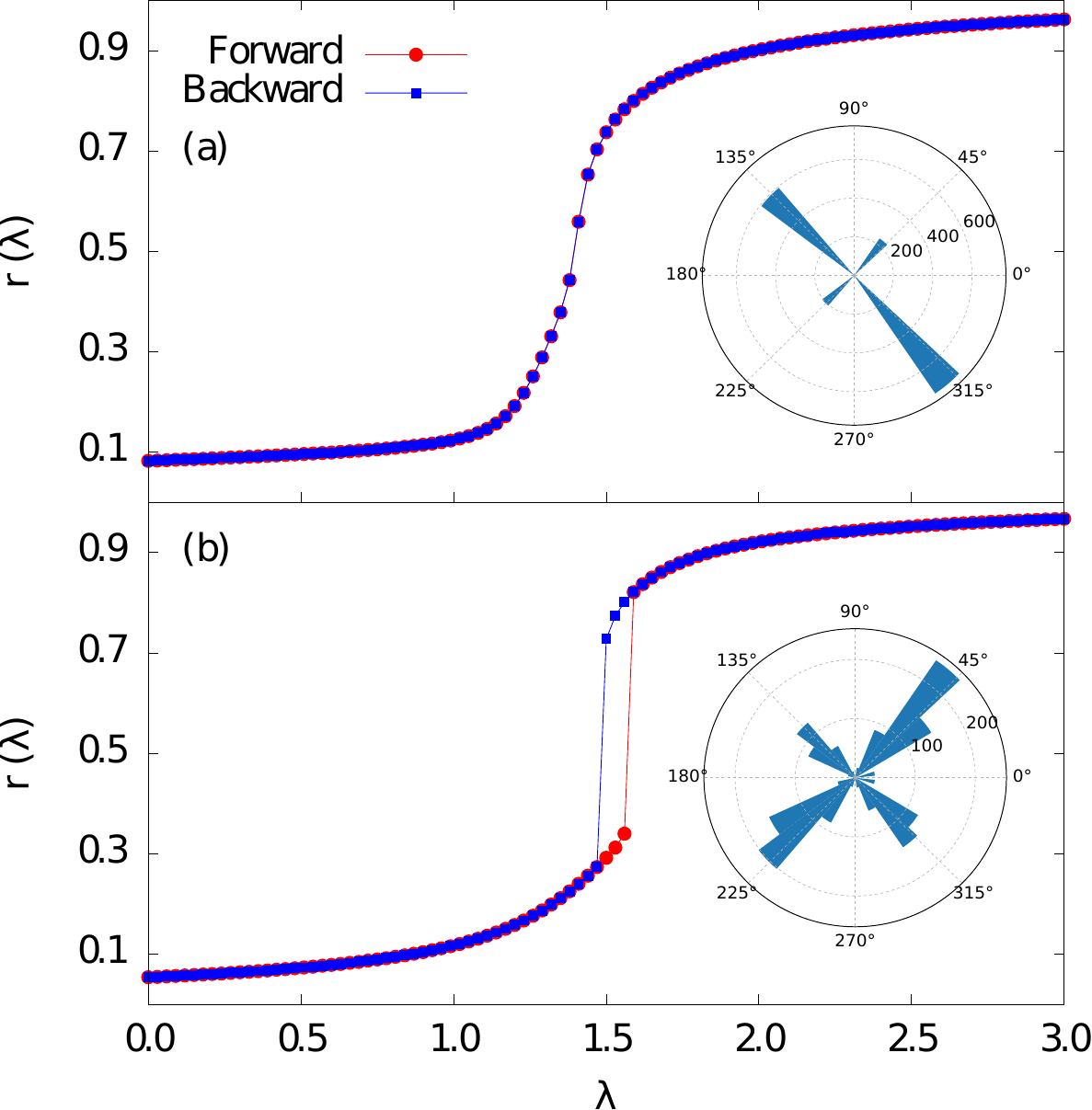}
    \caption{Main: Order parameter $r(\lambda)$ including both forward (red) and backward (blue) continuations in $\lambda$. 
    We consider correlated (a) and uncorrelated (b) empirical opinions from ANES data as $P(\varphi)$.
    Inset: Initial orientation distributions represented in polar coordinates at $\lambda = 0$. These empirical distributions are obtained neglecting all individuals with conviction lower than the median of $P(\rho)$, approaching then the case $\alpha \to \infty$. 
    }
    \label{fig:empirical_ordertransitions}
\end{figure}

Here we investigated how social influence can counter polarization, by proposing a simple analytically-tractable mean-field opinion dynamics model. 
In our model, an initial polarized state undergoes a depolarization transition to a consensus state, whose nature depends on the initial correlations between opinions: the depolarization is explosive (first order) when the initial opinions are uncorrelated. 
Our theoretical calculations are confirmed by numerical simulations based on real opinion patterns \add{as the initial polarization} collected from the ANES dataset. 
The model we propose represents a first step towards understanding the evolution of polarization in interdependent topics, in a very simple and intuitive setting. \delete{The model is versatile and it can be modified to take into account more realistic features, such as a pattern of social interactions given by a contact network, or considering the possible time evolution of conviction.}

\add{Our work, however, is not exempt from limitations. 
We considered a mean-field setting in which all individuals interact with everyone else. 
While this setting can be realistic for small interacting groups, interactions between agents are usually mediated by social networks: Future work should be dedicated to extending our analysis to networked substrates. 
Furthermore, we assumed the conviction of individuals to be constant in time, while it is reasonable that individuals change their conviction while changing orientation with respect to the two topics.} 
Finally, the \addd{social compass} model could be easily extended to $n$ topics, mapped in terms of spherical coordinates in $n$ dimensions, where the key assumption of the model (i.e., the stubborness of individuals proportional to their conviction) still holds. 

\vspace*{0.5cm}
\begin{acknowledgments}
  J.O. and R.P.S. acknowledge financial support from the Spanish
  MCIN/AEI/10.13039/501100011033, under Project No. PID2019-106290GB-C21.
\end{acknowledgments}

\bibliographystyle{apsrev4-1}
\bibliography{polaretti}{}

\begin{thebibliography}{48}%
\makeatletter
\providecommand \@ifxundefined [1]{%
 \@ifx{#1\undefined}
}%
\providecommand \@ifnum [1]{%
 \ifnum #1\expandafter \@firstoftwo
 \else \expandafter \@secondoftwo
 \fi
}%
\providecommand \@ifx [1]{%
 \ifx #1\expandafter \@firstoftwo
 \else \expandafter \@secondoftwo
 \fi
}%
\providecommand \natexlab [1]{#1}%
\providecommand \enquote  [1]{``#1''}%
\providecommand \bibnamefont  [1]{#1}%
\providecommand \bibfnamefont [1]{#1}%
\providecommand \citenamefont [1]{#1}%
\providecommand \href@noop [0]{\@secondoftwo}%
\providecommand \href [0]{\begingroup \@sanitize@url \@href}%
\providecommand \@href[1]{\@@startlink{#1}\@@href}%
\providecommand \@@href[1]{\endgroup#1\@@endlink}%
\providecommand \@sanitize@url [0]{\catcode `\\12\catcode `\$12\catcode
  `\&12\catcode `\#12\catcode `\^12\catcode `\_12\catcode `\%12\relax}%
\providecommand \@@startlink[1]{}%
\providecommand \@@endlink[0]{}%
\providecommand \url  [0]{\begingroup\@sanitize@url \@url }%
\providecommand \@url [1]{\endgroup\@href {#1}{\urlprefix }}%
\providecommand \urlprefix  [0]{URL }%
\providecommand \Eprint [0]{\href }%
\providecommand \doibase [0]{http://dx.doi.org/}%
\providecommand \selectlanguage [0]{\@gobble}%
\providecommand \bibinfo  [0]{\@secondoftwo}%
\providecommand \bibfield  [0]{\@secondoftwo}%
\providecommand \translation [1]{[#1]}%
\providecommand \BibitemOpen [0]{}%
\providecommand \bibitemStop [0]{}%
\providecommand \bibitemNoStop [0]{.\EOS\space}%
\providecommand \EOS [0]{\spacefactor3000\relax}%
\providecommand \BibitemShut  [1]{\csname bibitem#1\endcsname}%
\let\auto@bib@innerbib\@empty
\bibitem [{\citenamefont {Perry}(2022)}]{perry_american_2022}%
  \BibitemOpen
  \bibfield  {author} {\bibinfo {author} {\bibfnamefont {S.~L.}\ \bibnamefont
  {Perry}},\ }\href {\doibase 10.1146/annurev-soc-031021-114239} {\bibfield
  {journal} {\bibinfo  {journal} {Annual Review of Sociology}\ }\textbf
  {\bibinfo {volume} {48}},\ \bibinfo {pages} {87} (\bibinfo {year}
  {2022})}\BibitemShut {NoStop}%
\bibitem [{\citenamefont {Montalvo}\ and\ \citenamefont
  {Reynal-Querol}(2005)}]{montalvo_ethnic_2005}%
  \BibitemOpen
  \bibfield  {author} {\bibinfo {author} {\bibfnamefont {J.~G.}\ \bibnamefont
  {Montalvo}}\ and\ \bibinfo {author} {\bibfnamefont {M.}~\bibnamefont
  {Reynal-Querol}},\ }\href {\doibase 10.1257/0002828054201468} {\bibfield
  {journal} {\bibinfo  {journal} {American Economic Review}\ }\textbf {\bibinfo
  {volume} {95}},\ \bibinfo {pages} {796} (\bibinfo {year} {2005})}\BibitemShut
  {NoStop}%
\bibitem [{\citenamefont {McCright}\ and\ \citenamefont
  {Dunlap}(2011)}]{mccright2011politicization}%
  \BibitemOpen
  \bibfield  {author} {\bibinfo {author} {\bibfnamefont {A.~M.}\ \bibnamefont
  {McCright}}\ and\ \bibinfo {author} {\bibfnamefont {R.~E.}\ \bibnamefont
  {Dunlap}},\ }\href@noop {} {\bibfield  {journal} {\bibinfo  {journal} {The
  Sociological Quarterly}\ }\textbf {\bibinfo {volume} {52}},\ \bibinfo {pages}
  {155} (\bibinfo {year} {2011})}\BibitemShut {NoStop}%
\bibitem [{\citenamefont {McCoy}\ \emph {et~al.}(2018)\citenamefont {McCoy},
  \citenamefont {Rahman},\ and\ \citenamefont
  {Somer}}]{mccoy_polarization_2018}%
  \BibitemOpen
  \bibfield  {author} {\bibinfo {author} {\bibfnamefont {J.}~\bibnamefont
  {McCoy}}, \bibinfo {author} {\bibfnamefont {T.}~\bibnamefont {Rahman}}, \
  and\ \bibinfo {author} {\bibfnamefont {M.}~\bibnamefont {Somer}},\ }\href
  {\doibase 10.1177/0002764218759576} {\bibfield  {journal} {\bibinfo
  {journal} {American Behavioral Scientist}\ }\textbf {\bibinfo {volume}
  {62}},\ \bibinfo {pages} {16} (\bibinfo {year} {2018})}\BibitemShut {NoStop}%
\bibitem [{\citenamefont {Iyengar}\ \emph {et~al.}(2012)\citenamefont
  {Iyengar}, \citenamefont {Sood},\ and\ \citenamefont
  {Lelkes}}]{iyengar2012affect}%
  \BibitemOpen
  \bibfield  {author} {\bibinfo {author} {\bibfnamefont {S.}~\bibnamefont
  {Iyengar}}, \bibinfo {author} {\bibfnamefont {G.}~\bibnamefont {Sood}}, \
  and\ \bibinfo {author} {\bibfnamefont {Y.}~\bibnamefont {Lelkes}},\
  }\href@noop {} {\bibfield  {journal} {\bibinfo  {journal} {Public opinion
  quarterly}\ }\textbf {\bibinfo {volume} {76}},\ \bibinfo {pages} {405}
  (\bibinfo {year} {2012})}\BibitemShut {NoStop}%
\bibitem [{\citenamefont {Wang}\ \emph {et~al.}(2020)\citenamefont {Wang},
  \citenamefont {Jusup}, \citenamefont {Guo}, \citenamefont {Shi},
  \citenamefont {Ge{\v{c}}ek}, \citenamefont {Anand}, \citenamefont {Perc},
  \citenamefont {Bauch}, \citenamefont {Kurths}, \citenamefont {Boccaletti}
  \emph {et~al.}}]{wang2020communicating}%
  \BibitemOpen
  \bibfield  {author} {\bibinfo {author} {\bibfnamefont {Z.}~\bibnamefont
  {Wang}}, \bibinfo {author} {\bibfnamefont {M.}~\bibnamefont {Jusup}},
  \bibinfo {author} {\bibfnamefont {H.}~\bibnamefont {Guo}}, \bibinfo {author}
  {\bibfnamefont {L.}~\bibnamefont {Shi}}, \bibinfo {author} {\bibfnamefont
  {S.}~\bibnamefont {Ge{\v{c}}ek}}, \bibinfo {author} {\bibfnamefont
  {M.}~\bibnamefont {Anand}}, \bibinfo {author} {\bibfnamefont
  {M.}~\bibnamefont {Perc}}, \bibinfo {author} {\bibfnamefont {C.~T.}\
  \bibnamefont {Bauch}}, \bibinfo {author} {\bibfnamefont {J.}~\bibnamefont
  {Kurths}}, \bibinfo {author} {\bibfnamefont {S.}~\bibnamefont {Boccaletti}},
  \emph {et~al.},\ }\href@noop {} {\bibfield  {journal} {\bibinfo  {journal}
  {Proceedings of the National Academy of Sciences}\ }\textbf {\bibinfo
  {volume} {117}},\ \bibinfo {pages} {17650} (\bibinfo {year}
  {2020})}\BibitemShut {NoStop}%
\bibitem [{\citenamefont {Michael}(2017)}]{michael2017rise}%
  \BibitemOpen
  \bibfield  {author} {\bibinfo {author} {\bibfnamefont {G.}~\bibnamefont
  {Michael}},\ }\href@noop {} {\bibfield  {journal} {\bibinfo  {journal}
  {Skeptic (Altadena, CA)}\ }\textbf {\bibinfo {volume} {22}},\ \bibinfo
  {pages} {9} (\bibinfo {year} {2017})}\BibitemShut {NoStop}%
\bibitem [{\citenamefont {Dandekar}\ \emph {et~al.}(2013)\citenamefont
  {Dandekar}, \citenamefont {Goel},\ and\ \citenamefont
  {Lee}}]{dandekar_biased_2013}%
  \BibitemOpen
  \bibfield  {author} {\bibinfo {author} {\bibfnamefont {P.}~\bibnamefont
  {Dandekar}}, \bibinfo {author} {\bibfnamefont {A.}~\bibnamefont {Goel}}, \
  and\ \bibinfo {author} {\bibfnamefont {D.~T.}\ \bibnamefont {Lee}},\ }\href
  {\doibase 10.1073/pnas.1217220110} {\bibfield  {journal} {\bibinfo  {journal}
  {Proceedings of the National Academy of Sciences}\ }\textbf {\bibinfo
  {volume} {110}},\ \bibinfo {pages} {5791} (\bibinfo {year}
  {2013})}\BibitemShut {NoStop}%
\bibitem [{\citenamefont {Baumann}\ \emph {et~al.}(2020)\citenamefont
  {Baumann}, \citenamefont {Lorenz-Spreen}, \citenamefont {Sokolov},\ and\
  \citenamefont {Starnini}}]{baumann19}%
  \BibitemOpen
  \bibfield  {author} {\bibinfo {author} {\bibfnamefont {F.}~\bibnamefont
  {Baumann}}, \bibinfo {author} {\bibfnamefont {P.}~\bibnamefont
  {Lorenz-Spreen}}, \bibinfo {author} {\bibfnamefont {I.~M.}\ \bibnamefont
  {Sokolov}}, \ and\ \bibinfo {author} {\bibfnamefont {M.}~\bibnamefont
  {Starnini}},\ }\href {\doibase 10.1103/PhysRevLett.124.048301} {\bibfield
  {journal} {\bibinfo  {journal} {Phys. Rev. Lett.}\ }\textbf {\bibinfo
  {volume} {124}},\ \bibinfo {pages} {048301} (\bibinfo {year}
  {2020})}\BibitemShut {NoStop}%
\bibitem [{\citenamefont {Deffuant}\ \emph {et~al.}(2000)\citenamefont
  {Deffuant}, \citenamefont {Neau}, \citenamefont {Amblard},\ and\
  \citenamefont {Weisbuch}}]{deffuant_mixing_2000}%
  \BibitemOpen
  \bibfield  {author} {\bibinfo {author} {\bibfnamefont {G.}~\bibnamefont
  {Deffuant}}, \bibinfo {author} {\bibfnamefont {D.}~\bibnamefont {Neau}},
  \bibinfo {author} {\bibfnamefont {F.}~\bibnamefont {Amblard}}, \ and\
  \bibinfo {author} {\bibfnamefont {G.}~\bibnamefont {Weisbuch}},\ }\href
  {\doibase 10.1142/S0219525900000078} {\bibfield  {journal} {\bibinfo
  {journal} {Advs. Complex Syst.}\ }\textbf {\bibinfo {volume} {03}},\ \bibinfo
  {pages} {87} (\bibinfo {year} {2000})}\BibitemShut {NoStop}%
\bibitem [{\citenamefont {Hegselmann}\ and\ \citenamefont
  {Krause}(2002)}]{hegselmann_opinion_2002}%
  \BibitemOpen
  \bibfield  {author} {\bibinfo {author} {\bibfnamefont {R.}~\bibnamefont
  {Hegselmann}}\ and\ \bibinfo {author} {\bibfnamefont {U.}~\bibnamefont
  {Krause}},\ }\href@noop {} {\bibfield  {journal} {\bibinfo  {journal}
  {Journal of Artificial Societies and Social Simulation}\ }\textbf {\bibinfo
  {volume} {5}},\ \bibinfo {pages} {1} (\bibinfo {year} {2002})}\BibitemShut
  {NoStop}%
\bibitem [{\citenamefont {Lorenz}(2007)}]{lorenz_continuous_2007}%
  \BibitemOpen
  \bibfield  {author} {\bibinfo {author} {\bibfnamefont {J.}~\bibnamefont
  {Lorenz}},\ }\href {\doibase 10.1142/S0129183107011789} {\bibfield  {journal}
  {\bibinfo  {journal} {Int. J. Mod. Phys. C}\ }\textbf {\bibinfo {volume}
  {18}},\ \bibinfo {pages} {1819} (\bibinfo {year} {2007})}\BibitemShut
  {NoStop}%
\bibitem [{\citenamefont {Huet}\ \emph {et~al.}(2008)\citenamefont {Huet},
  \citenamefont {Deffuant},\ and\ \citenamefont {Jager}}]{huet_rejection_2008}%
  \BibitemOpen
  \bibfield  {author} {\bibinfo {author} {\bibfnamefont {S.}~\bibnamefont
  {Huet}}, \bibinfo {author} {\bibfnamefont {G.}~\bibnamefont {Deffuant}}, \
  and\ \bibinfo {author} {\bibfnamefont {W.}~\bibnamefont {Jager}},\ }\href
  {\doibase 10.1142/S0219525908001799} {\bibfield  {journal} {\bibinfo
  {journal} {Advs. Complex Syst.}\ }\textbf {\bibinfo {volume} {11}},\ \bibinfo
  {pages} {529} (\bibinfo {year} {2008})}\BibitemShut {NoStop}%
\bibitem [{\citenamefont {Crawford}\ \emph {et~al.}(2013)\citenamefont
  {Crawford}, \citenamefont {Brooks},\ and\ \citenamefont
  {Sen}}]{crawford_opposites_2013}%
  \BibitemOpen
  \bibfield  {author} {\bibinfo {author} {\bibfnamefont {C.}~\bibnamefont
  {Crawford}}, \bibinfo {author} {\bibfnamefont {L.}~\bibnamefont {Brooks}}, \
  and\ \bibinfo {author} {\bibfnamefont {S.}~\bibnamefont {Sen}},\ }in\
  \href@noop {} {\emph {\bibinfo {booktitle} {Proceedings of the 2013
  international conference on {Autonomous} agents and multi-agent systems}}},\
  \bibinfo {series and number} {{AAMAS} '13}\ (\bibinfo  {publisher}
  {International Foundation for Autonomous Agents and Multiagent Systems},\
  \bibinfo {address} {Richland, SC},\ \bibinfo {year} {2013})\ pp.\ \bibinfo
  {pages} {1225--1226}\BibitemShut {NoStop}%
\bibitem [{\citenamefont {Vinokur}\ and\ \citenamefont
  {Burnstein}(1978)}]{vinokur1978depolarization}%
  \BibitemOpen
  \bibfield  {author} {\bibinfo {author} {\bibfnamefont {A.}~\bibnamefont
  {Vinokur}}\ and\ \bibinfo {author} {\bibfnamefont {E.}~\bibnamefont
  {Burnstein}},\ }\href@noop {} {\bibfield  {journal} {\bibinfo  {journal}
  {Journal of Personality and Social Psychology}\ }\textbf {\bibinfo {volume}
  {36}},\ \bibinfo {pages} {872} (\bibinfo {year} {1978})}\BibitemShut
  {NoStop}%
\bibitem [{\citenamefont {Matakos}\ \emph {et~al.}(2017)\citenamefont
  {Matakos}, \citenamefont {Terzi},\ and\ \citenamefont
  {Tsaparas}}]{matakos_measuring_2017}%
  \BibitemOpen
  \bibfield  {author} {\bibinfo {author} {\bibfnamefont {A.}~\bibnamefont
  {Matakos}}, \bibinfo {author} {\bibfnamefont {E.}~\bibnamefont {Terzi}}, \
  and\ \bibinfo {author} {\bibfnamefont {P.}~\bibnamefont {Tsaparas}},\ }\href
  {\doibase 10.1007/s10618-017-0527-9} {\bibfield  {journal} {\bibinfo
  {journal} {Data Min Knowl Disc}\ }\textbf {\bibinfo {volume} {31}},\ \bibinfo
  {pages} {1480} (\bibinfo {year} {2017})}\BibitemShut {NoStop}%
\bibitem [{\citenamefont {Musco}\ \emph {et~al.}(2018)\citenamefont {Musco},
  \citenamefont {Musco},\ and\ \citenamefont
  {Tsourakakis}}]{musco_minimizing_2018}%
  \BibitemOpen
  \bibfield  {author} {\bibinfo {author} {\bibfnamefont {C.}~\bibnamefont
  {Musco}}, \bibinfo {author} {\bibfnamefont {C.}~\bibnamefont {Musco}}, \ and\
  \bibinfo {author} {\bibfnamefont {C.~E.}\ \bibnamefont {Tsourakakis}},\ }in\
  \href {\doibase 10.1145/3178876.3186103} {\emph {\bibinfo {booktitle}
  {Proceedings of the 2018 {World} {Wide} {Web} {Conference}}}},\ \bibinfo
  {series and number} {{WWW} '18}\ (\bibinfo  {publisher} {International World
  Wide Web Conferences Steering Committee},\ \bibinfo {address} {Republic and
  Canton of Geneva, CHE},\ \bibinfo {year} {2018})\ pp.\ \bibinfo {pages}
  {369--378}\BibitemShut {NoStop}%
\bibitem [{\citenamefont {Balietti}\ \emph {et~al.}(2021)\citenamefont
  {Balietti}, \citenamefont {Getoor}, \citenamefont {Goldstein},\ and\
  \citenamefont {Watts}}]{balietti_reducing_2021}%
  \BibitemOpen
  \bibfield  {author} {\bibinfo {author} {\bibfnamefont {S.}~\bibnamefont
  {Balietti}}, \bibinfo {author} {\bibfnamefont {L.}~\bibnamefont {Getoor}},
  \bibinfo {author} {\bibfnamefont {D.~G.}\ \bibnamefont {Goldstein}}, \ and\
  \bibinfo {author} {\bibfnamefont {D.~J.}\ \bibnamefont {Watts}},\ }\href
  {\doibase 10.1073/pnas.2112552118} {\bibfield  {journal} {\bibinfo  {journal}
  {Proceedings of the National Academy of Sciences}\ }\textbf {\bibinfo
  {volume} {118}},\ \bibinfo {pages} {e2112552118} (\bibinfo {year}
  {2021})}\BibitemShut {NoStop}%
\bibitem [{\citenamefont {Jager}\ and\ \citenamefont
  {Amblard}(2005)}]{jager_uniformity_2005}%
  \BibitemOpen
  \bibfield  {author} {\bibinfo {author} {\bibfnamefont {W.}~\bibnamefont
  {Jager}}\ and\ \bibinfo {author} {\bibfnamefont {F.}~\bibnamefont
  {Amblard}},\ }\href {\doibase 10.1007/s10588-005-6282-2} {\bibfield
  {journal} {\bibinfo  {journal} {Comput Math Organiz Theor}\ }\textbf
  {\bibinfo {volume} {10}},\ \bibinfo {pages} {295} (\bibinfo {year}
  {2005})}\BibitemShut {NoStop}%
\bibitem [{\citenamefont {Chau}\ \emph {et~al.}(2014)\citenamefont {Chau},
  \citenamefont {Wong}, \citenamefont {Chow},\ and\ \citenamefont
  {Fung}}]{chau_social_2014}%
  \BibitemOpen
  \bibfield  {author} {\bibinfo {author} {\bibfnamefont {H.~F.}\ \bibnamefont
  {Chau}}, \bibinfo {author} {\bibfnamefont {C.~Y.}\ \bibnamefont {Wong}},
  \bibinfo {author} {\bibfnamefont {F.~K.}\ \bibnamefont {Chow}}, \ and\
  \bibinfo {author} {\bibfnamefont {C.-H.~F.}\ \bibnamefont {Fung}},\ }\href
  {\doibase 10.1016/j.physa.2014.07.082} {\bibfield  {journal} {\bibinfo
  {journal} {Physica A: Statistical Mechanics and its Applications}\ }\textbf
  {\bibinfo {volume} {415}},\ \bibinfo {pages} {133} (\bibinfo {year}
  {2014})}\BibitemShut {NoStop}%
\bibitem [{\citenamefont {Poole}(2005)}]{poole_spatial_2005}%
  \BibitemOpen
  \bibfield  {author} {\bibinfo {author} {\bibfnamefont {K.~T.}\ \bibnamefont
  {Poole}},\ }\href {\doibase 10.1017/CBO9780511614644} {\emph {\bibinfo
  {title} {Spatial {Models} of {Parliamentary} {Voting}}}},\ Analytical
  {Methods} for {Social} {Research}\ (\bibinfo {address} {Cambridge},\ \bibinfo
  {year} {2005})\BibitemShut {NoStop}%
\bibitem [{\citenamefont {Benoit}\ and\ \citenamefont
  {Laver}(2012)}]{benoit_dimensionality_2012}%
  \BibitemOpen
  \bibfield  {author} {\bibinfo {author} {\bibfnamefont {K.}~\bibnamefont
  {Benoit}}\ and\ \bibinfo {author} {\bibfnamefont {M.}~\bibnamefont {Laver}},\
  }\href {\doibase 10.1177/1465116511434618} {\bibfield  {journal} {\bibinfo
  {journal} {European Union Politics}\ }\textbf {\bibinfo {volume} {13}},\
  \bibinfo {pages} {194} (\bibinfo {year} {2012})}\BibitemShut {NoStop}%
\bibitem [{\citenamefont {Li}\ and\ \citenamefont
  {Xiao}(2017)}]{li_agent-based_2017}%
  \BibitemOpen
  \bibfield  {author} {\bibinfo {author} {\bibfnamefont {J.}~\bibnamefont
  {Li}}\ and\ \bibinfo {author} {\bibfnamefont {R.}~\bibnamefont {Xiao}},\
  }\href@noop {} {\bibfield  {journal} {\bibinfo  {journal} {JASSS}\ }\textbf
  {\bibinfo {volume} {20}},\ \bibinfo {pages} {4} (\bibinfo {year}
  {2017})}\BibitemShut {NoStop}%
\bibitem [{\citenamefont {van~der Maas}\ \emph {et~al.}(2020)\citenamefont
  {van~der Maas}, \citenamefont {Dalege},\ and\ \citenamefont
  {Waldorp}}]{van_der_maas_polarization_2020}%
  \BibitemOpen
  \bibfield  {author} {\bibinfo {author} {\bibfnamefont {H.~L.~J.}\
  \bibnamefont {van~der Maas}}, \bibinfo {author} {\bibfnamefont
  {J.}~\bibnamefont {Dalege}}, \ and\ \bibinfo {author} {\bibfnamefont
  {L.}~\bibnamefont {Waldorp}},\ }\href {\doibase 10.1093/comnet/cnaa010}
  {\bibfield  {journal} {\bibinfo  {journal} {Journal of Complex Networks}\
  }\textbf {\bibinfo {volume} {8}},\ \bibinfo {pages} {cnaa010} (\bibinfo
  {year} {2020})}\BibitemShut {NoStop}%
\bibitem [{\citenamefont {Schweighofer}\ \emph {et~al.}(2020)\citenamefont
  {Schweighofer}, \citenamefont {Garcia},\ and\ \citenamefont
  {Schweitzer}}]{schweighofer_agent-based_2020}%
  \BibitemOpen
  \bibfield  {author} {\bibinfo {author} {\bibfnamefont {S.}~\bibnamefont
  {Schweighofer}}, \bibinfo {author} {\bibfnamefont {D.}~\bibnamefont
  {Garcia}}, \ and\ \bibinfo {author} {\bibfnamefont {F.}~\bibnamefont
  {Schweitzer}},\ }\href {\doibase 10.1063/5.0007523} {\bibfield  {journal}
  {\bibinfo  {journal} {Chaos}\ }\textbf {\bibinfo {volume} {30}},\ \bibinfo
  {pages} {093139} (\bibinfo {year} {2020})}\BibitemShut {NoStop}%
\bibitem [{\citenamefont {Chen}\ \emph {et~al.}(2021)\citenamefont {Chen},
  \citenamefont {Wang}, \citenamefont {Yang},\ and\ \citenamefont
  {Cong}}]{chen_modeling_2021}%
  \BibitemOpen
  \bibfield  {author} {\bibinfo {author} {\bibfnamefont {T.}~\bibnamefont
  {Chen}}, \bibinfo {author} {\bibfnamefont {Y.}~\bibnamefont {Wang}}, \bibinfo
  {author} {\bibfnamefont {J.}~\bibnamefont {Yang}}, \ and\ \bibinfo {author}
  {\bibfnamefont {G.}~\bibnamefont {Cong}},\ }\href {\doibase
  10.3390/ijerph18020472} {\bibfield  {journal} {\bibinfo  {journal}
  {International Journal of Environmental Research and Public Health}\ }\textbf
  {\bibinfo {volume} {18}},\ \bibinfo {pages} {472} (\bibinfo {year}
  {2021})}\BibitemShut {NoStop}%
\bibitem [{\citenamefont {DiMaggio}\ \emph {et~al.}(1996)\citenamefont
  {DiMaggio}, \citenamefont {Evans},\ and\ \citenamefont
  {Bryson}}]{dimaggio_have_1996}%
  \BibitemOpen
  \bibfield  {author} {\bibinfo {author} {\bibfnamefont {P.}~\bibnamefont
  {DiMaggio}}, \bibinfo {author} {\bibfnamefont {J.}~\bibnamefont {Evans}}, \
  and\ \bibinfo {author} {\bibfnamefont {B.}~\bibnamefont {Bryson}},\ }\href
  {\doibase 10.1086/230995} {\bibfield  {journal} {\bibinfo  {journal}
  {American Journal of Sociology}\ }\textbf {\bibinfo {volume} {102}},\
  \bibinfo {pages} {690} (\bibinfo {year} {1996})}\BibitemShut {NoStop}%
\bibitem [{\citenamefont {Baldassarri}\ and\ \citenamefont
  {Gelman}(2008)}]{baldassarri_partisans_2008}%
  \BibitemOpen
  \bibfield  {author} {\bibinfo {author} {\bibfnamefont {D.}~\bibnamefont
  {Baldassarri}}\ and\ \bibinfo {author} {\bibfnamefont {A.}~\bibnamefont
  {Gelman}},\ }\href {\doibase 10.2139/ssrn.1010098} {\bibfield  {journal}
  {\bibinfo  {journal} {AJS}\ }\textbf {\bibinfo {volume} {114}},\ \bibinfo
  {pages} {408} (\bibinfo {year} {2008})}\BibitemShut {NoStop}%
\bibitem [{\citenamefont {Freire}(2008)}]{freire_party_2008}%
  \BibitemOpen
  \bibfield  {author} {\bibinfo {author} {\bibfnamefont {A.}~\bibnamefont
  {Freire}},\ }\href {\doibase 10.1177/1354068807085889} {\bibfield  {journal}
  {\bibinfo  {journal} {Party Politics}\ }\textbf {\bibinfo {volume} {14}},\
  \bibinfo {pages} {189} (\bibinfo {year} {2008})}\BibitemShut {NoStop}%
\bibitem [{\citenamefont {Falck}\ \emph {et~al.}(2020)\citenamefont {Falck},
  \citenamefont {Marstaller}, \citenamefont {Stoehr}, \citenamefont {Maucher},
  \citenamefont {Ren}, \citenamefont {Thalhammer}, \citenamefont {Rettinger},\
  and\ \citenamefont {Studer}}]{falck_measuring_2020}%
  \BibitemOpen
  \bibfield  {author} {\bibinfo {author} {\bibfnamefont {F.}~\bibnamefont
  {Falck}}, \bibinfo {author} {\bibfnamefont {J.}~\bibnamefont {Marstaller}},
  \bibinfo {author} {\bibfnamefont {N.}~\bibnamefont {Stoehr}}, \bibinfo
  {author} {\bibfnamefont {S.}~\bibnamefont {Maucher}}, \bibinfo {author}
  {\bibfnamefont {J.}~\bibnamefont {Ren}}, \bibinfo {author} {\bibfnamefont
  {A.}~\bibnamefont {Thalhammer}}, \bibinfo {author} {\bibfnamefont
  {A.}~\bibnamefont {Rettinger}}, \ and\ \bibinfo {author} {\bibfnamefont
  {R.}~\bibnamefont {Studer}},\ }\href {\doibase 10.1002/poi3.222} {\bibfield
  {journal} {\bibinfo  {journal} {Policy \& Internet}\ }\textbf {\bibinfo
  {volume} {12}},\ \bibinfo {pages} {367} (\bibinfo {year} {2020})}\BibitemShut
  {NoStop}%
\bibitem [{\citenamefont {Adamczyk}(2022)}]{adamczyk_religion_2022}%
  \BibitemOpen
  \bibfield  {author} {\bibinfo {author} {\bibfnamefont {A.}~\bibnamefont
  {Adamczyk}},\ }\href {\doibase 10.1093/esr/jcac017} {\bibfield  {journal}
  {\bibinfo  {journal} {European Sociological Review}\ ,\ \bibinfo {pages}
  {jcac017}} (\bibinfo {year} {2022})}\BibitemShut {NoStop}%
\bibitem [{\citenamefont {DellaPosta}\ \emph {et~al.}(2015)\citenamefont
  {DellaPosta}, \citenamefont {Shi},\ and\ \citenamefont
  {Macy}}]{dellaposta_why_2015}%
  \BibitemOpen
  \bibfield  {author} {\bibinfo {author} {\bibfnamefont {D.}~\bibnamefont
  {DellaPosta}}, \bibinfo {author} {\bibfnamefont {Y.}~\bibnamefont {Shi}}, \
  and\ \bibinfo {author} {\bibfnamefont {M.}~\bibnamefont {Macy}},\ }\href
  {\doibase 10.1086/681254} {\bibfield  {journal} {\bibinfo  {journal}
  {American Journal of Sociology}\ }\textbf {\bibinfo {volume} {120}},\
  \bibinfo {pages} {1473} (\bibinfo {year} {2015})}\BibitemShut {NoStop}%
\bibitem [{\citenamefont {Laguna}\ \emph {et~al.}(2003)\citenamefont {Laguna},
  \citenamefont {Abramson},\ and\ \citenamefont
  {Zanette}}]{laguna_vector_2003}%
  \BibitemOpen
  \bibfield  {author} {\bibinfo {author} {\bibfnamefont {M.~F.}\ \bibnamefont
  {Laguna}}, \bibinfo {author} {\bibfnamefont {G.}~\bibnamefont {Abramson}}, \
  and\ \bibinfo {author} {\bibfnamefont {D.~H.}\ \bibnamefont {Zanette}},\
  }\href {\doibase 10.1016/S0378-4371(03)00628-9} {\bibfield  {journal}
  {\bibinfo  {journal} {Physica A: Statistical Mechanics and its Applications}\
  }\textbf {\bibinfo {volume} {329}},\ \bibinfo {pages} {459} (\bibinfo {year}
  {2003})}\BibitemShut {NoStop}%
\bibitem [{\citenamefont {Fortunato}\ \emph {et~al.}(2005)\citenamefont
  {Fortunato}, \citenamefont {Latora}, \citenamefont {Pluchino},\ and\
  \citenamefont {Rapisarda}}]{fortunato_vector_2005}%
  \BibitemOpen
  \bibfield  {author} {\bibinfo {author} {\bibfnamefont {S.}~\bibnamefont
  {Fortunato}}, \bibinfo {author} {\bibfnamefont {V.}~\bibnamefont {Latora}},
  \bibinfo {author} {\bibfnamefont {A.}~\bibnamefont {Pluchino}}, \ and\
  \bibinfo {author} {\bibfnamefont {A.}~\bibnamefont {Rapisarda}},\ }\href
  {\doibase 10.1142/S0129183105008126} {\bibfield  {journal} {\bibinfo
  {journal} {Int. J. Mod. Phys. C}\ }\textbf {\bibinfo {volume} {16}},\
  \bibinfo {pages} {1535} (\bibinfo {year} {2005})}\BibitemShut {NoStop}%
\bibitem [{\citenamefont {Etesami}\ \emph {et~al.}(2013)\citenamefont
  {Etesami}, \citenamefont {Başar}, \citenamefont {Nedić},\ and\
  \citenamefont {Touri}}]{etesami_termination_2013}%
  \BibitemOpen
  \bibfield  {author} {\bibinfo {author} {\bibfnamefont {S.~R.}\ \bibnamefont
  {Etesami}}, \bibinfo {author} {\bibfnamefont {T.}~\bibnamefont {Başar}},
  \bibinfo {author} {\bibfnamefont {A.}~\bibnamefont {Nedić}}, \ and\ \bibinfo
  {author} {\bibfnamefont {B.}~\bibnamefont {Touri}},\ }in\ \href {\doibase
  10.1109/ACC.2013.6580008} {\emph {\bibinfo {booktitle} {2013 {American}
  {Control} {Conference}}}}\ (\bibinfo {year} {2013})\ pp.\ \bibinfo {pages}
  {1255--1260},\ \bibinfo {note} {iSSN: 2378-5861}\BibitemShut {NoStop}%
\bibitem [{\citenamefont {Converse}(2006)}]{converse_nature_2006}%
  \BibitemOpen
  \bibfield  {author} {\bibinfo {author} {\bibfnamefont {P.~E.}\ \bibnamefont
  {Converse}},\ }\href {\doibase 10.1080/08913810608443650} {\bibfield
  {journal} {\bibinfo  {journal} {Critical Review}\ }\textbf {\bibinfo {volume}
  {18}},\ \bibinfo {pages} {1} (\bibinfo {year} {2006})}\BibitemShut {NoStop}%
\bibitem [{\citenamefont {Baumann}\ \emph {et~al.}(2021)\citenamefont
  {Baumann}, \citenamefont {Lorenz-Spreen}, \citenamefont {Sokolov},\ and\
  \citenamefont {Starnini}}]{baumann_emergence_2021}%
  \BibitemOpen
  \bibfield  {author} {\bibinfo {author} {\bibfnamefont {F.}~\bibnamefont
  {Baumann}}, \bibinfo {author} {\bibfnamefont {P.}~\bibnamefont
  {Lorenz-Spreen}}, \bibinfo {author} {\bibfnamefont {I.~M.}\ \bibnamefont
  {Sokolov}}, \ and\ \bibinfo {author} {\bibfnamefont {M.}~\bibnamefont
  {Starnini}},\ }\href {\doibase 10.1103/PhysRevX.11.011012} {\bibfield
  {journal} {\bibinfo  {journal} {Phys. Rev. X}\ }\textbf {\bibinfo {volume}
  {11}},\ \bibinfo {pages} {011012} (\bibinfo {year} {2021})}\BibitemShut
  {NoStop}%
\bibitem [{\citenamefont {Abelson}(1988)}]{abelson_conviction_1988}%
  \BibitemOpen
  \bibfield  {author} {\bibinfo {author} {\bibfnamefont {R.~P.}\ \bibnamefont
  {Abelson}},\ }\href {\doibase 10.1037/0003-066X.43.4.267} {\bibfield
  {journal} {\bibinfo  {journal} {American Psychologist}\ }\textbf {\bibinfo
  {volume} {43}},\ \bibinfo {pages} {267} (\bibinfo {year} {1988})}\BibitemShut
  {NoStop}%
\bibitem [{\citenamefont {Miller}\ \emph {et~al.}(1993)\citenamefont {Miller},
  \citenamefont {McHoskey}, \citenamefont {Bane},\ and\ \citenamefont
  {Dowd}}]{miller_attitude_1993}%
  \BibitemOpen
  \bibfield  {author} {\bibinfo {author} {\bibfnamefont {A.~G.}\ \bibnamefont
  {Miller}}, \bibinfo {author} {\bibfnamefont {J.~W.}\ \bibnamefont
  {McHoskey}}, \bibinfo {author} {\bibfnamefont {C.~M.}\ \bibnamefont {Bane}},
  \ and\ \bibinfo {author} {\bibfnamefont {T.~G.}\ \bibnamefont {Dowd}},\
  }\href {\doibase 10.1037/0022-3514.64.4.561} {\bibfield  {journal} {\bibinfo
  {journal} {Journal of Personality and Social Psychology}\ }\textbf {\bibinfo
  {volume} {64}},\ \bibinfo {pages} {561} (\bibinfo {year} {1993})}\BibitemShut
  {NoStop}%
\bibitem [{\citenamefont {Pomerantz}\ \emph {et~al.}(1995)\citenamefont
  {Pomerantz}, \citenamefont {Chaiken},\ and\ \citenamefont
  {Tordesillas}}]{pomerantz_attitude_1995}%
  \BibitemOpen
  \bibfield  {author} {\bibinfo {author} {\bibfnamefont {E.~M.}\ \bibnamefont
  {Pomerantz}}, \bibinfo {author} {\bibfnamefont {S.}~\bibnamefont {Chaiken}},
  \ and\ \bibinfo {author} {\bibfnamefont {R.~S.}\ \bibnamefont
  {Tordesillas}},\ }\href {\doibase 10.1037/0022-3514.69.3.408} {\bibfield
  {journal} {\bibinfo  {journal} {Journal of Personality and Social
  Psychology}\ }\textbf {\bibinfo {volume} {69}},\ \bibinfo {pages} {408}
  (\bibinfo {year} {1995})}\BibitemShut {NoStop}%
\bibitem [{Note1()}]{Note1}%
  \BibitemOpen
  \bibinfo {note} {\protect \url {http://www.politicalcompass.org}}\BibitemShut
  {NoStop}%
\bibitem [{Note2()}]{Note2}%
  \BibitemOpen
  \bibinfo {note} {\protect \url {http://www.electionstudies.org}}\BibitemShut
  {NoStop}%
\bibitem [{\citenamefont {Friedkin}\ and\ \citenamefont
  {Johnsen}(1990)}]{friedkin_social_1990}%
  \BibitemOpen
  \bibfield  {author} {\bibinfo {author} {\bibfnamefont {N.~E.}\ \bibnamefont
  {Friedkin}}\ and\ \bibinfo {author} {\bibfnamefont {E.~C.}\ \bibnamefont
  {Johnsen}},\ }\href {\doibase 10.1080/0022250X.1990.9990069} {\bibfield
  {journal} {\bibinfo  {journal} {The Journal of Mathematical Sociology}\
  }\textbf {\bibinfo {volume} {15}},\ \bibinfo {pages} {193} (\bibinfo {year}
  {1990})}\BibitemShut {NoStop}%
\bibitem [{\citenamefont {Pluchino}\ \emph {et~al.}(2005)\citenamefont
  {Pluchino}, \citenamefont {Latora},\ and\ \citenamefont
  {Rapisarda}}]{pluchino_changing_2005}%
  \BibitemOpen
  \bibfield  {author} {\bibinfo {author} {\bibfnamefont {A.}~\bibnamefont
  {Pluchino}}, \bibinfo {author} {\bibfnamefont {V.}~\bibnamefont {Latora}}, \
  and\ \bibinfo {author} {\bibfnamefont {A.}~\bibnamefont {Rapisarda}},\ }\href
  {\doibase 10.1142/S0129183105007261} {\bibfield  {journal} {\bibinfo
  {journal} {Int. J. Mod. Phys. C}\ }\textbf {\bibinfo {volume} {16}},\
  \bibinfo {pages} {515} (\bibinfo {year} {2005})}\BibitemShut {NoStop}%
\bibitem [{\citenamefont {Castellano}\ \emph {et~al.}(2009)\citenamefont
  {Castellano}, \citenamefont {Fortunato},\ and\ \citenamefont
  {Loreto}}]{castellano_statistical_2009}%
  \BibitemOpen
  \bibfield  {author} {\bibinfo {author} {\bibfnamefont {C.}~\bibnamefont
  {Castellano}}, \bibinfo {author} {\bibfnamefont {S.}~\bibnamefont
  {Fortunato}}, \ and\ \bibinfo {author} {\bibfnamefont {V.}~\bibnamefont
  {Loreto}},\ }\href {\doibase 10.1103/RevModPhys.81.591} {\bibfield  {journal}
  {\bibinfo  {journal} {Rev. Mod. Phys.}\ }\textbf {\bibinfo {volume} {81}},\
  \bibinfo {pages} {591} (\bibinfo {year} {2009})}\BibitemShut {NoStop}%
\bibitem [{\citenamefont {Kuramoto}(1975)}]{kuramoto_self-entrainment_1975}%
  \BibitemOpen
  \bibfield  {author} {\bibinfo {author} {\bibfnamefont {Y.}~\bibnamefont
  {Kuramoto}},\ }in\ \href {\doibase 10.1007/BFb0013365} {\emph {\bibinfo
  {booktitle} {International {Symposium} on {Mathematical} {Problems} in
  {Theoretical} {Physics}}}},\ \bibinfo {series and number} {Lecture {Notes} in
  {Physics}},\ \bibinfo {editor} {edited by\ \bibinfo {editor} {\bibfnamefont
  {H.}~\bibnamefont {Araki}}}\ (\bibinfo  {publisher} {Springer},\ \bibinfo
  {address} {Berlin, Heidelberg},\ \bibinfo {year} {1975})\ pp.\ \bibinfo
  {pages} {420--422}\BibitemShut {NoStop}%
\bibitem [{\citenamefont {Acebrón}\ \emph {et~al.}(2005)\citenamefont
  {Acebrón}, \citenamefont {Bonilla}, \citenamefont {Vicente}, \citenamefont
  {Ritort},\ and\ \citenamefont {Spigler}}]{Acebron.2005}%
  \BibitemOpen
  \bibfield  {author} {\bibinfo {author} {\bibfnamefont {J.~A.}\ \bibnamefont
  {Acebrón}}, \bibinfo {author} {\bibfnamefont {L.~L.}\ \bibnamefont
  {Bonilla}}, \bibinfo {author} {\bibfnamefont {C.~J.~P.}\ \bibnamefont
  {Vicente}}, \bibinfo {author} {\bibfnamefont {F.}~\bibnamefont {Ritort}}, \
  and\ \bibinfo {author} {\bibfnamefont {R.}~\bibnamefont {Spigler}},\ }\href
  {\doibase 10.1103/revmodphys.77.137} {\bibfield  {journal} {\bibinfo
  {journal} {Reviews of Modern Physics}\ }\textbf {\bibinfo {volume} {77}},\
  \bibinfo {pages} {137} (\bibinfo {year} {2005})}\BibitemShut {NoStop}%
\bibitem [{\citenamefont {Abramowitz}\ and\ \citenamefont
  {Stegun}(1972)}]{abramovitz}%
  \BibitemOpen
  \bibfield  {author} {\bibinfo {author} {\bibfnamefont {M.}~\bibnamefont
  {Abramowitz}}\ and\ \bibinfo {author} {\bibfnamefont {I.~A.}\ \bibnamefont
  {Stegun}},\ }\href@noop {} {\emph {\bibinfo {title} {Handbook of mathematical
  functions.}}}\ (\bibinfo  {publisher} {Dover},\ \bibinfo {address} {New
  York},\ \bibinfo {year} {1972})\BibitemShut {NoStop}%
\end{thebibliography}%

\pagebreak

\newcommand{\thisisthetitle}{Modeling Explosive Opinion Depolarization in Interdependent Topics}

\pagestyle{fancy}

\lhead{\thisisthetitle\ - SM}
\chead{}
\rhead{}

\lfoot{J. Ojer, M. Starnini and R. Pastor-Satorras}
\cfoot{}
\rfoot{\thepage}

\renewcommand{\headrulewidth}{0.5pt}
\renewcommand{\footrulewidth}{0.5pt}

\renewcommand{\thefigure}{\textbf{SF~\arabic{figure}}}
\setcounter{figure}{0}
\renewcommand{\figurename}{\textbf{Supplementary Figure}}
\setcounter{equation}{0}

\title{\thisisthetitle\ \\ \ \\
  Supplementary Material}

\author{Jaume Ojer}

\affiliation{Departament de F\'{\i}sica, Universitat Polit\`ecnica de
  Catalunya, Campus Nord B4, 08034 Barcelona, Spain}

\author{Michele Starnini}

\affiliation{Departament de F\'{\i}sica, Universitat Polit\`ecnica de
  Catalunya, Campus Nord B4, 08034 Barcelona, Spain}

\author{Romualdo Pastor-Satorras}

\affiliation{Departament de F\'{\i}sica, Universitat Polit\`ecnica de
  Catalunya, Campus Nord B4, 08034 Barcelona, Spain}

\maketitle
\onecolumngrid
\begin{center}
\textbf{\large Supplementary Material}
\end{center}

\section{A polar projection of two-dimensional opinions}
\label{sec:polar_projection}

The ANES study is a continuation of a series of academically-run surveys since 1948, asking several questions to a representative sample of citizens in the United States with the objective of analyzing public opinion and voting behavior during the presidential elections. In order to test our two-dimensional model, we selected 12 different pairs combining 12 distinct questions with 49380 valid responses in total from the 2016 ANES. The questions have been chosen with the following criteria. 
First, since the analysis of the empirical dataset has to be scaled numerically, we focused on multiple-choice questions, excluding questions with free-text answers like ``What is your main occupation?''. 
Second, we selected questions with a scale of a minimum of three points, 
in order to quantify the conviction with an adequate intensity range.
This allows us to extract the sufficient extent of approval or disapproval (favor, neutral or oppose) of the respondent regarding the particular topic.

Then, we focused on two classes of topics: polarizing and non-polarizing. 
With respect to polarizing topics,  we chose questions where most respondents hold extreme opinions, such as "favor strongly" or "oppose a great deal". 
This will ensure a confronting scenario in the two-dimensional topic space far from an opinion consensus. 
Here we itemize by an issue label the different ANES questions used throughout the work, detailing the possible answers as a numerical point scale:

\begin{enumerate}
\item \texttt{religion}: Does religion provide you guidance in day-to-day living? $1$ some, $2$ quite a bit, $3$ a great deal. ANES ID: V161242.
\item \texttt{same sex}: Do you think business owners who provide wedding-related services should be allowed to refuse services to same-sex couples if same-sex marriage violates their religious beliefs? agree ($1$ strongly, $2$ moderately, $3$ a little), disagree ($4$ a little, $5$ moderately, $6$ strongly). ANES ID: V161227x.
\item \texttt{obamacare}: What do you think about 2010 health care law? favor ($1$ a great deal, $2$ moderately, $3$ a little), $4$ neutral, oppose ($5$ a little, $6$ moderately, $7$ a great deal). ANES ID: V161114x.
\item \texttt{transgender}: Should transgender people have to use the bathrooms of the gender they were born as? agree ($1$ strongly, $2$ moderately, $3$ a little), disagree ($4$ a little, $5$ moderately, $6$ strongly). ANES ID: V161228x.
\item \texttt{birthright}: Should the U.S. Constitution be changed so that the children of unauthorized immigrants do not automatically get citizenship if they are born in this country? favor ($1$ a great deal, $2$ moderately, $3$ a little), $4$ neutral, oppose ($5$ a little, $6$ moderately, $7$ a great deal). ANES ID: V161194x.
\item \texttt{fight ISIS}: Should U.S. send ground troops to fight Islamic militants, such as ISIS, in Iraq and Syria? favor ($1$ a great deal, $2$ moderately, $3$ a little), $4$ neutral, oppose ($5$ a little, $6$ moderately, $7$ a great deal). ANES ID: V161213x.
\item \texttt{Mexican wall}: What do you think about building a wall on the U.S. border with Mexico? favor ($1$ a great deal, $2$ moderately, $3$ a little), $4$ neutral, oppose ($5$ a little, $6$ moderately, $7$ a great deal). ANES ID: V161196x.
\item \texttt{climate change}: Is the federal government doing the right amount about rising temperatures? should be more ($1$ a great deal, $2$ moderately, $3$ a little), $4$ right amount, should be less ($5$ a little, $6$ moderately, $7$ a great deal). ANES ID: V161225x.
\end{enumerate}
As not all the questions share the same answer scale, we normalized all the scales to the range $[-1, 1]$.


However, not all of the different pairs resulting from combinations of the distinct questions show a sufficiently low order parameter $r$ (see Eq.~(2) in the main paper). 
Since we are interested in initially polarized states, we chose the following pairs with order parameter $r < 0.2$:
\begin{enumerate}
\item \texttt{obamacare} and \texttt{transgender}.
\item \texttt{obamacare} and \texttt{same sex}.
\item \texttt{obamacare} and \texttt{religion}.
\item \texttt{religion} and \texttt{same sex}.
\item \texttt{religion} and \texttt{fight ISIS}.
\item \texttt{transgender} and \texttt{religion}.
\item \texttt{transgender} and \texttt{same sex}.
\item \texttt{Mexican wall} and \texttt{same sex}.
\item \texttt{climate change} and \texttt{fight ISIS}.
\item \texttt{birthright} and \texttt{fight ISIS}.
\end{enumerate}

With respect to non-polarizing  topics, we chose the following questions:
\begin{enumerate}
\item \texttt{leader}: Is a strong leader in government good for U.S. even if the leader bends rules to get things done? agree ($1$ strongly, $2$ somewhat), $3$ neutral, disagree ($4$ somewhat, $5$ strongly). ANES ID: V162263.
\item \texttt{born in U.S.}: How important to have been born in U.S. is for being truly American? $1$ a lot, $2$ fairly, $3$ not much, $4$ not at all. ANES ID: V162271.
\item \texttt{feminism}: How well does the term ``feminist" describe you? $1$ extremely, $2$ very well, $3$ somewhat, $4$ not very well, $5$ not at all. ANES ID: V161346.
\item \texttt{women \& men}: Do women fail to appreciate fully all that men do for them? agree ($1$ strongly, $2$ somewhat), $3$ neutral, disagree ($4$ somewhat, $5$ strongly). ANES ID: V161508.
\end{enumerate}
In the following, we consider the pairs:

\begin{enumerate}
\item \texttt{leader} and \texttt{born in U.S.}
\item \texttt{feminism} and \texttt{women \& men}.
\end{enumerate}

\subsection{Angular distribution}
\label{sec:polar_ang}

\begin{figure}[tbp]
    \centering
    \includegraphics[width=0.5\textwidth]{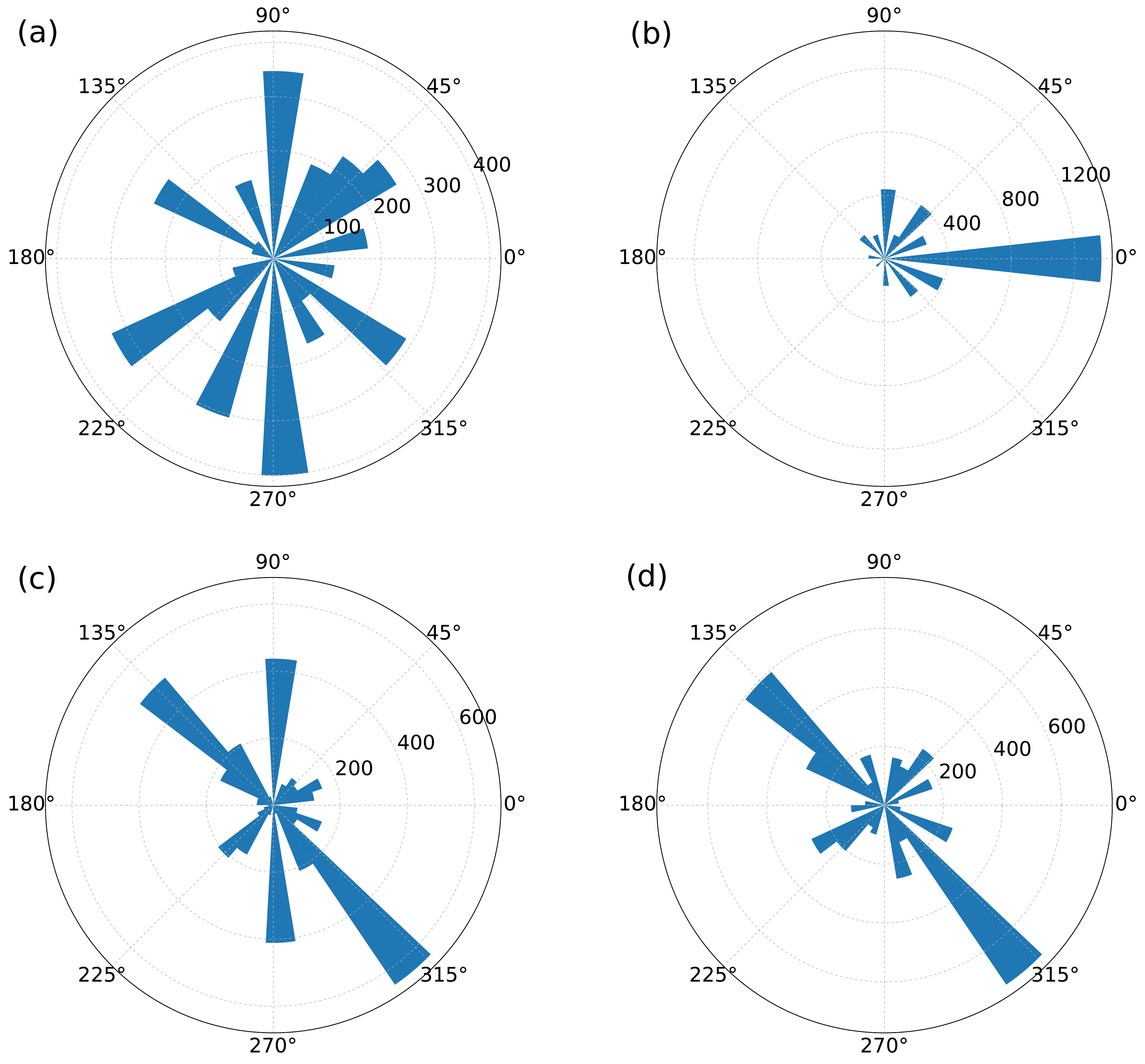}
    \caption{Empirical ANES opinions about two topics represented as histograms in a polar projection. Different distributions are obtained depending on the issue combination: a) \texttt{leader} and \texttt{born in U.S.}, uniform (order parameter $r = 0.079$), b) \texttt{feminism} and \texttt{women \& men}, consensus ($r = 0.543$), c) \texttt{obamacare} and \texttt{transgender}, uncorrelated polarization ($r = 0.067$), and d) \texttt{religion} and \texttt{same sex}, correlated polarization ($r = 0.030$). 
    }
    \label{fig:ANES_polar}
\end{figure}

In \ref{fig:ANES_polar} we show four different interesting cases of the possible distributions of opinions regarding two real topics in the polar plane. If opinions are more or less uniformly distributed, there is no peak around any specific opinion, so the angular variable $\varphi$ will also be roughly uniformly distributed, see \ref{fig:ANES_polar}(a). If instead there is a consensus around both topics, the angular distribution $P(\varphi)$ will be peaked around a certain value $\varphi^\ast = \arctan(y^\ast/x^\ast)$, where $y^\ast$ and $x^\ast$ are respectively the consensus opinions of topics $Y$ and $X$, see \ref{fig:ANES_polar}(b). Then, if the opinions with respect to both topics are polarized, i.e., the opinion distribution in one dimension is characterized by the presence of two well separated peaks, two different cases are possible. If there is no correlation between opinions regarding topic $X$ and topic $Y$, then the $P(\varphi)$ distribution will be characterized by four peaks, as shown in \ref{fig:ANES_polar}(c). If instead opinions with respect to both topics are correlated, then one will only observe two peaks in the $P(\varphi)$, see \ref{fig:ANES_polar}(d).

\subsection{Conviction distribution}
\label{sec:polar_conv}

\begin{figure}[h!]
    \centering
    \includegraphics[width=0.5\textwidth]{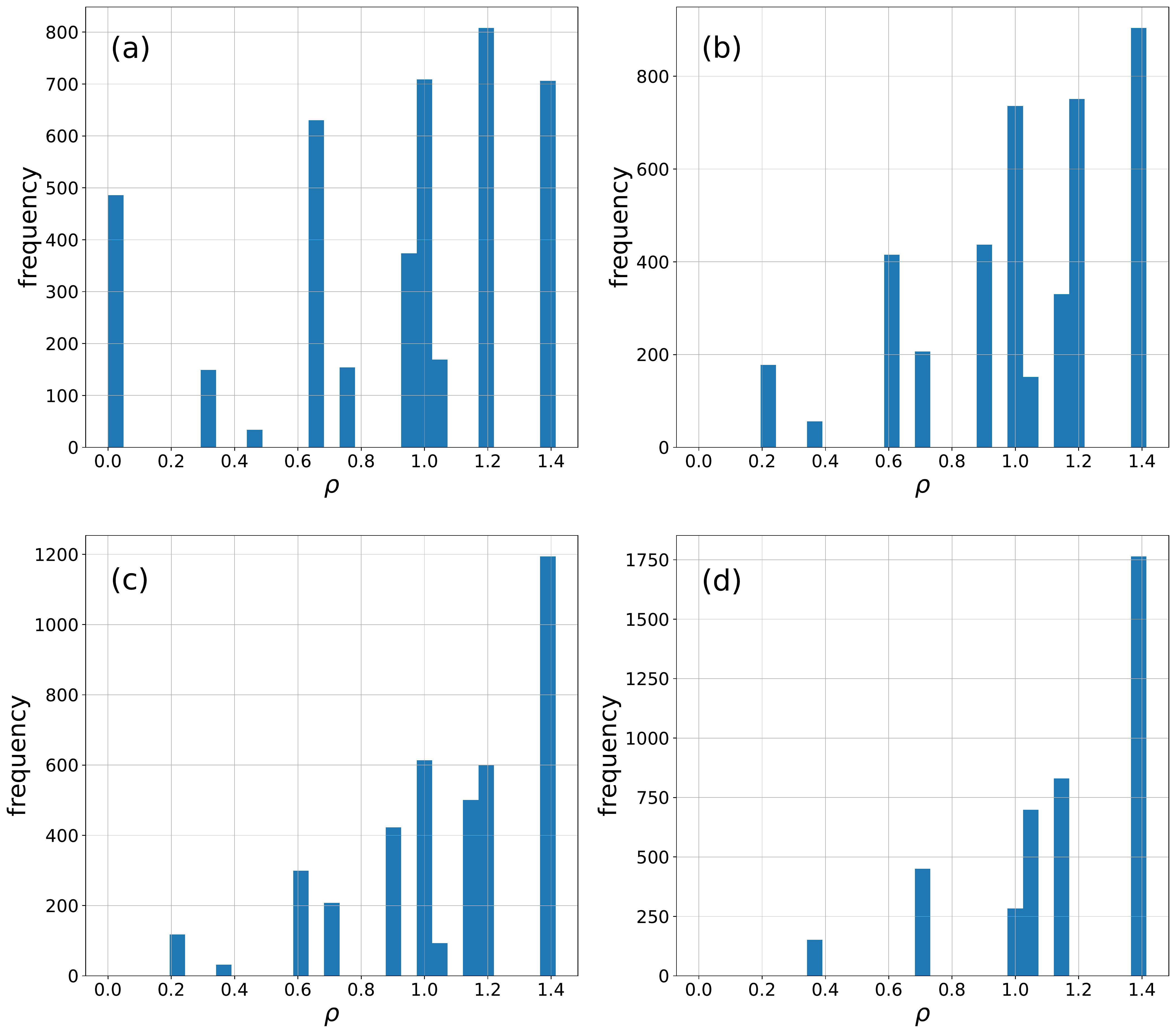}
    \caption{Empirical ANES convictions about two topics represented as histograms. Different dependencies as $P(\rho) \sim \rho^\alpha$ can be related: a) \texttt{birthright} and \texttt{fight ISIS}, uniform ($\alpha \simeq 0$), b) \texttt{fight ISIS} and \texttt{same sex}, growing as $\rho$ ($\alpha \simeq 1$), c) \texttt{obamacare} and \texttt{transgender}, growing as a power of $\rho$ ($\alpha > 1$), and d) \texttt{religion} and \texttt{same sex}, the majority holding maximum conviction ($\alpha \gg 1$).}
    \label{fig:ANES_rho}
\end{figure}

In \ref{fig:ANES_rho} we show different profiles of the conviction distribution for pairs of distinct real topics. As we can see, the conviction distribution is in general either rather constant or a growing function of $\rho$. As a quantitative  pattern of this distribution we thus propose the general approximation $P(\rho) \sim \rho^\alpha$, where $\alpha=0$ represents a uniform distribution while $\alpha > 0$ represents a growing distribution with different levels of heterogeneity.


\section{General solution of the model}
\label{sec:general_solution}

The mean-field character of the model allows us to decouple each individual opinion from all the others through the mean-field quantities $r$ and $\psi$.
Following Ref.~\cite{Acebron.2005}, from the equation for the order parameter
\begin{equation}
    r(\lambda)\mathrm{e}^{i\psi(\lambda)} = \frac{1}{N} \sum_{j = 1}^N \mathrm{e}^{i\theta_j(\lambda)},
    \label{eq:order_parameter}
\end{equation}
multiplying both sides by $\mathrm{e}^{-i\theta_i(\lambda)}$ and taking the imaginary part, we obtain
\begin{equation}
    r(\lambda)\sin{[\psi(\lambda) - \theta_i(\lambda)]} = \frac{1}{N} \sum_{j = 1}^N \sin{[\theta_j(\lambda) - \theta_i(\lambda)]}.
    \label{eq:order_MF}
\end{equation}
Replacing this expression into Eq.~(1) of the main text, we obtain
\begin{equation}
    \dot{\theta}_i = \rho_i \sin(\varphi_i - \theta_i) + \lambda r \sin(\psi - \theta_i),
    \label{eq:model_MF}
\end{equation}
a set of equations that are only coupled by the common average angle $\psi$, and where we have omitted the implicit dependence on $\lambda$ by $\theta_i$, $r$ and $\psi$ for simplicity. In the steady state, setting the time derivative equal to zero, we can obtain explicitly the fixed point $\bar{\theta_i}$ for any values of the parameters. Indeed, developing both angle differences and dividing all by $\cos{(\theta_i)}$, we find
\begin{equation}
\rho_i \left[ \sin{(\varphi_i)} - \cos{(\varphi_i)}\tan{(\bar{\theta_i})} \right] + \lambda r \left[ \sin{(\psi)} - \cos{(\psi)}\tan{(\bar{\theta_i})} \right] = 0, \nonumber
\end{equation}
which finally leads to 
\begin{equation}
    \bar{\theta}_i(\lambda, \rho, \varphi) = \arctan \left( \frac{\lambda r \sin(\psi) + \rho_i \sin(\varphi_i)}{\lambda r \cos(\psi) + \rho_i \cos(\varphi_i)} \right).
\label{eq:fixed_point}
\end{equation}
The stationary regime is described in the thermodynamic limit by the probability $p(\theta | \rho, \varphi)$ that an agent with a conviction $\rho$ and an initial opinion $\varphi$, which are distributed according to given uncorrelated probability densities $P(\rho)$ and $P(\varphi)$, respectively, shows an orientation $\theta$. Therefore, the order parameter can be rewritten in the continuum limit as
\begin{equation}
r(\lambda) \mathrm{e}^{i\psi(\lambda)} = \int_0^\infty d\rho \int_{-\pi}^\pi d\varphi \int_{-\pi}^\pi d\theta \, p(\theta | \rho, \varphi) P(\rho) P(\varphi) \, \mathrm{e}^{i\theta(\lambda)}.
\label{eq:continuum}
\end{equation}
Hence, since it is assumed that a fixed point is reached at sufficiently large times, $\theta$ is given deterministically by Eq.~\eqref{eq:fixed_point}, so the stationary distribution is defined as $p(\theta | \rho, \varphi) = \delta(\theta - \bar{\theta})$. Inserting and rearranging it in Eq.~\eqref{eq:continuum}, we obtain
\begin{equation}
r(\lambda) = \int_0^\infty d\rho \int_{-\pi}^\pi d\varphi P(\rho) P(\varphi) \, \mathrm{e}^{i \left[\bar{\theta}(\lambda, \rho, \varphi) - \psi(\lambda)\right]},
\label{eq:general}
\end{equation}
a self-consistent relation between the mean-field quantities $r$ and $\psi$ and the model variables $\lambda$, $\rho$ and $\varphi$. Then, we can separate the real part from the imaginary one of the form
\begin{eqnarray}
r(\lambda) = \int_0^\infty d\rho \int_{-\pi}^{\pi} d\varphi P(\rho) P(\varphi) \cos\left[ \bar{\theta}(\lambda, \rho, \varphi) - \psi(\lambda) \right] \nonumber \\
0 = \int_0^\infty d\rho \int_{-\pi}^{\pi} d\varphi P(\rho) P(\varphi) \sin\left[ \bar{\theta}(\lambda, \rho, \varphi) - \psi(\lambda) \right]. \nonumber
\end{eqnarray}
Developing the angle differences and using the identities $\cos{\left[ \arctan(x) \right]} = 1/\sqrt{1 + x^2}$ and $\sin{\left[ \arctan(x) \right]} = x/\sqrt{1 + x^2}$, the equations extend to
\begin{eqnarray}
r &=& \int_0^\infty d\rho P(\rho) \int_{-\pi}^{\pi} d\varphi P(\varphi) \frac{\left[ \lambda r \cos{(\psi)} + \rho \cos(\varphi) \right] \cos{(\psi)}}{\sqrt{\left[ \lambda r \cos{(\psi)} + \rho \cos(\varphi) \right]^2 + \left[ \lambda r \sin{(\psi)} + \rho \sin(\varphi) \right]^2}} \nonumber \\
&+& \int_0^\infty d\rho P(\rho) \int_{-\pi}^{\pi} d\varphi P(\varphi) \frac{\left[ \lambda r \sin{(\psi)} + \rho \sin(\varphi) \right] \sin{(\psi)}}{\sqrt{\left[ \lambda r \cos{(\psi)} + \rho \cos(\varphi) \right]^2 + \left[ \lambda r \sin{(\psi)} + \rho \sin(\varphi) \right]^2}}, \nonumber  \\
0 &=& \int_0^\infty d\rho P(\rho) \int_{-\pi}^{\pi} d\varphi P(\varphi) \frac{\left[ \lambda r \sin{(\psi)} + \rho \sin(\varphi) \right] \cos{(\psi)}}{\sqrt{\left[ \lambda r \cos{(\psi)} + \rho \cos(\varphi) \right]^2 + \left[ \lambda r \sin{(\psi)} + \rho \sin(\varphi) \right]^2}} \nonumber \\
&-& \int_0^\infty d\rho P(\rho) \int_{-\pi}^{\pi} d\varphi P(\varphi) \frac{\left[ \lambda r \cos{(\psi)} + \rho \cos(\varphi) \right] \sin{(\psi)}}{\sqrt{\left[ \lambda r \cos{(\psi)} + \rho \cos(\varphi) \right]^2 + \left[ \lambda r \sin{(\psi)} + \rho \sin(\varphi) \right]^2}}. \nonumber
\end{eqnarray}
After some algebra, we finally get
\begin{eqnarray}
r = \int_0^\infty d\rho P(\rho) \int_{-\pi}^{\pi} d\varphi P(\varphi) \frac{\lambda r + \rho \cos(\varphi - \psi)}{\sqrt{(\lambda r)^2 + 2\lambda r \rho \cos(\varphi - \psi ) + \rho^2}}, \label{eq:real_part}  \\
0 = \int_0^\infty d\rho P(\rho) \int_{-\pi}^{\pi} d\varphi P(\varphi) \frac{\rho \sin(\varphi - \psi)}{\sqrt{(\lambda r)^2 + 2\lambda r \rho \cos(\varphi - \psi) + \rho^2}}, \label{eq:imaginary_part}
\end{eqnarray}
which can be reduced to
\begin{equation}
r = \int_0^\infty d\rho \int_{-\pi}^{\pi} d\varphi \frac{P(\rho) P(\varphi) \left[ \lambda r + \rho \, \mathrm{e}^{i \left( \varphi - \psi \right)} \right]}{\sqrt{(\lambda r)^2 + 2\lambda r \rho \cos(\varphi - \psi ) + \rho^2}}, \label{eq:general}
\end{equation}
corresponding to Eq.~(3) of the main text.

\subsection{Critical point}
\label{sec:general_critical}

Defining $r = I(r, \psi)$ from Eq.~\eqref{eq:general} (Eq.~(3) of the main text), the threshold condition can be obtained imposing instability for the solution $r = 0$, that is
\begin{equation}
\begin{aligned}
\left. \frac{\partial \mathrm{Re} \{I(r, \psi) \}}{\partial r} \right|_{r = 0} 
& = \int_0^\infty d\rho \int_{-\pi}^{\pi} d\varphi P(\rho) P(\varphi)  \left[ \frac{\lambda}{\left[ (\lambda r)^2 + 2\lambda r \rho \cos{(\varphi - \psi)} + \rho^2 \right]^{1/2}} \right. \\
                                    & \left. \left. - \enspace \frac{\left[ 2 \lambda^2 r + 2 \lambda \rho \cos{(\varphi - \psi)} \right] \left[ \lambda r + \rho \cos{(\varphi - \psi)} \right]}{2\left[ (\lambda r)^2 + 2\lambda r \rho \cos{(\varphi - \psi)} + \rho^2 \right]^{3/2}} \right] \right|_{r = 0} \\
                                    & = \int_0^\infty d\rho \int_{-\pi}^{\pi} d\varphi P(\rho) P(\varphi) \frac{\lambda}{\rho} \sin^2(\varphi - \psi) \ge 1, \nonumber
\end{aligned}
\end{equation}
where ``$\mathrm{Re}$'' stands for the real part. The critical point for the onset of instability 
 reads as
\begin{equation}
\lambda_c = \frac{1}{\int_0^\infty d\rho \frac{P(\rho)}{\rho} \int_{-\pi}^{\pi}d\varphi P(\varphi) \sin^2(\varphi - \psi)},
\label{eq:critical_point}
\end{equation}
which corresponds to Eq.~(4) of the main text.

\subsection{Empirical distributions}
\label{sec:general_empirical}

\begin{figure}[tbp]
     \centering

     \includegraphics[width=0.27\textwidth]{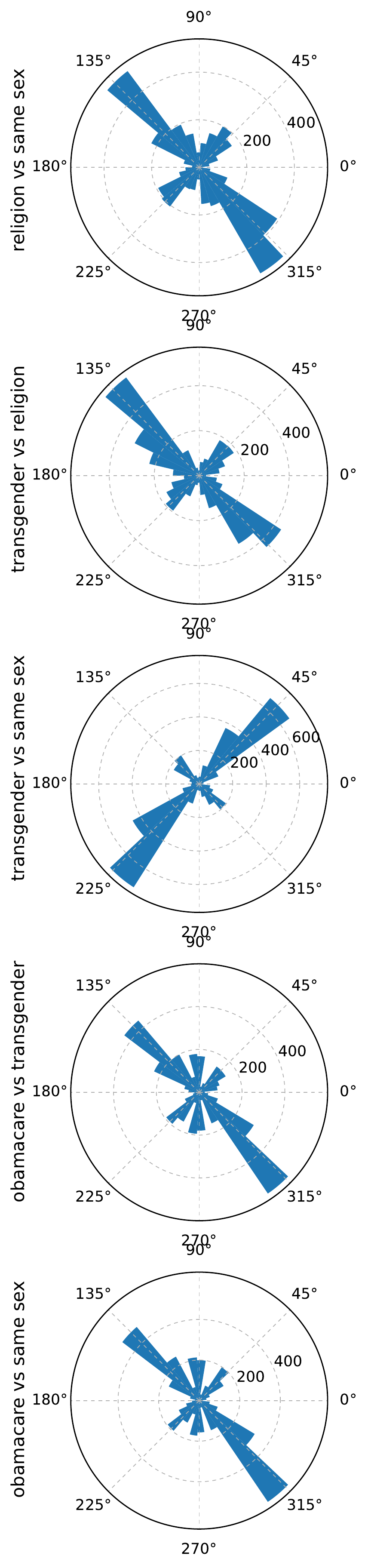}
     \includegraphics[width=0.27\textwidth]{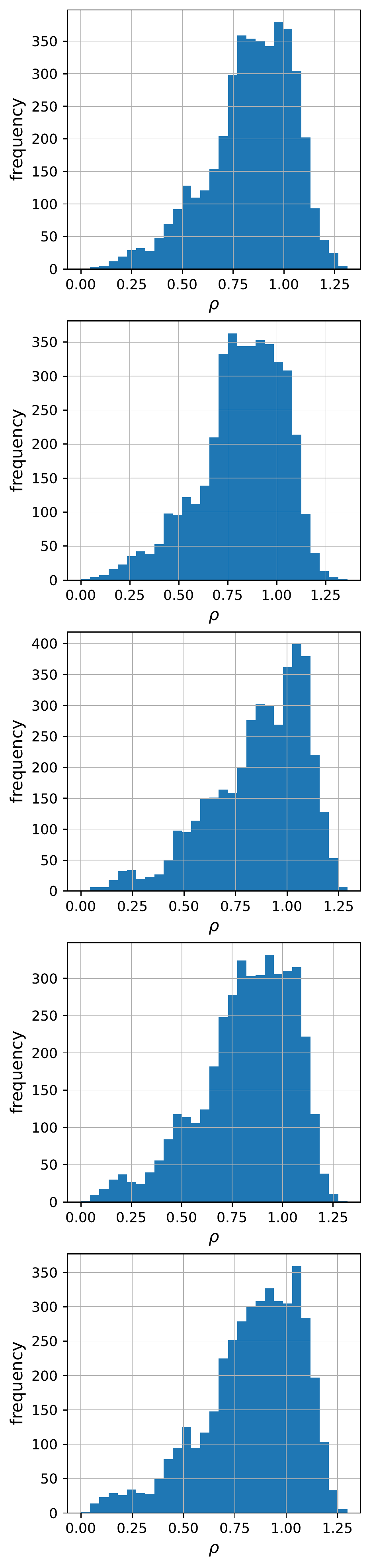}
        \caption{Empirical ANES opinions in a polar projection (left, approximately bimodal distributions) and convictions (right) about two topics used in Fig.~1 of the main text. 
        }
        \label{fig:empirical}
\end{figure}

We tested the validity of the theoretical prediction for the critical point (Fig.~1 of the main text) by using 5 empirical ANES distributions with approximately a bimodal initial opinion distribution (see \ref{fig:empirical}). Since the conviction distributions $P(\rho)$ are discrete rather than continuous (see \ref{fig:ANES_rho}), we slightly jittered empirical data by applying a small noise.
Due to this jittering procedure, one can now observe a peak in the conviction distribution, which however is just an artifact and does not change the true functional form of the original distribution.

Following Eq.~\eqref{eq:critical_point} (Eq.~(4) of the main text), the numerical computation of the critical point is done as
\begin{equation}
\lambda_c^\textrm{num} = \frac{1}{\av{\frac{1}{\rho}}_\rho \av{\sin^2(\varphi - \psi)}_\varphi},
\label{eq:critical_num}
\end{equation}
where $\av{...}$ denotes the average value of the numerical data. 
For correlated polarization, the average orientation $\psi$ is perpendicular to the direction of the angular orientation of the bimodal distribution $\phi_0$, $\psi = \phi_0 \pm \pi/2$ (see Sec.~\ref{sec:correlated_bimodal}, Eq.~\eqref{eq:bimodal_psi}).
Thus, for the empirical distributions, we qualitatively chose an average orientation $\psi$ perpendicular to $\phi_0$. 
For example, for the first case of \ref{fig:empirical} considering ``religion'' and ``same sex'', we set $\psi = 40 \degree$.

\subsection{Average orientation}
\label{sec:general_average}

In general, the value of $\psi$ can be inferred from Eq.~\eqref{eq:imaginary_part} for given probability densities $P(\rho)$ and $P(\varphi)$. We see, however, that the conviction is positive by definition, so its definite integral may not contribute to set the equation equal to zero. The only requirement that has to be fulfilled in order to satisfy the previous assumption is that the integral
\begin{equation}
\Theta (\rho) = \int_{-\pi}^{\pi} d\varphi P(\varphi) \frac{\sin(\varphi - \psi)}{\sqrt{(\lambda r)^2 + 2\lambda r \rho \cos(\varphi - \psi) + \rho^2}} \label{eq:angular_integral}
\end{equation}
can not change sign as $\rho$ varies within its positive domain. If this is met, then the average orientation $\psi$ can be determined from $\Theta = 0$, depending only on $P(\varphi)$ and regardless of $P(\rho)$. 
In the following sections, we investigate it in more detail for bimodal and quadrimodal angular distributions.

\section{Solution for correlated polarization: bimodal distribution}
\label{sec:correlated_bimodal}

A bimodal distribution with two antiparallel peaks is given by Eq.~(6) of the main text with $\varphi_0 = 0$. 
Here we consider a more general form where $\phi_0\in [0, \pi)$ is the angular orientation with respect to the $x$-axis
\begin{equation}
P(\varphi) = \frac{1}{2} \delta(\varphi - \phi_0) + \frac{1}{2} \delta(\varphi - \phi_0 + \pi).
\label{eq:bimodal}
\end{equation}
Its integration in Eq.~\eqref{eq:angular_integral} gives rise to
\begin{equation}
\Theta (\rho) = \frac{\sin(\phi_0 - \psi)}{2} \left[ \frac{1}{\sqrt{(\lambda r)^2 + 2\lambda r \rho \cos(\phi_0 - \psi) + \rho^2}} - \frac{1}{\sqrt{(\lambda r)^2 - 2\lambda r \rho \cos(\phi_0 - \psi) + \rho^2}} \right]. \label{eq:bimodal_integral}
\end{equation}
The first factor of the right-hand side does not depend on $\rho$, whilst the second factor is always positive (if $\cos(\phi_0 - \psi) < 0$), negative (if $\cos(\phi_0 - \psi) > 0$) or zero (if $\cos(\phi_0 - \psi) = 0$) for any value of $\rho, \lambda \in [0, \infty]$. For a correlated polarization, therefore, the average orientation is independent of the conviction and the coupling constant.

Now we can infer the value of $\psi$ by means of the condition $\Theta = 0$. 
From Eq.~\eqref{eq:bimodal_integral} it is easy to see that there are two possibilities. 
The first one is $\sin{(\phi_0 - \psi)} = 0$, which means $\psi = \phi_0$ or $\psi = \phi_0 \pm \pi$, i.e., an average orientation aligned with one of the two peaks of the distribution. This is physically impossible, since replacing it in Eq.~\eqref{eq:real_part} we obtain $r = 1$ for any value of $\lambda$, which is in contradiction with the definition of polarization ($r = 0$ would not be a solution). The other possibility is that the second factor vanishes with $\cos(\phi_0 - \psi) = -\cos(\phi_0 - \psi)$, i.e.,
\begin{equation}
\psi = \phi_0 \pm \pi/2.
\label{eq:bimodal_psi}
\end{equation}
Although this is a general result valid for any $\phi_0$, in the main text we set $\phi_0 = 0$ for simplicity, which corresponds to $\psi = \pm \pi/2$.

\subsection{Power-law distributed conviction}
\label{sec:correlated_power}

In general, for a power-law conviction distribution with $\alpha > 0$ and $\rho_\mathrm{max} = 1$, the real part of the self-consistent function $I(r, \psi)$ extracted from Eq.~\eqref{eq:real_part} reads as
\begin{equation}
\mathrm{Re} \{ I(r, \psi) \} = (\alpha + 1) \int_0^1 d\rho \frac{\lambda r \rho^\alpha}{\sqrt{(\lambda r)^2 + \rho^2}} = \, _2F_1 \left( \frac{1}{2}, \frac{\alpha + 1}{2}; \frac{\alpha + 3}{2}; -\frac{1}{(\lambda r)^2} \right),
\label{eq:self_general_supl}
\end{equation}
where $_2F_1(a,b;c;z)$ is the Gaussian hypergeometric function. We can study the behavior in the vicinity of the phase transition by expanding the hypergeometric function and studying its asymptotic behavior at $r \to 0$. To do so, we use
\begin{eqnarray}
_2F_1 (a,b;c;z) &\simeq& \frac{\Gamma(b - a)\Gamma(c)}{\Gamma(b)\Gamma(c - a)}(-z)^{-a} \left[ 1 + \frac{a(1 + a - c)}{(1 + a - b)z} + \mathcal{O}(\frac{1}{z^2}) \right] \nonumber \\
&+& \frac{\Gamma(a - b)\Gamma(c)}{\Gamma(a)\Gamma(c - b)}(-z)^{-b} \left[ 1 + \frac{b(1 + b - c)}{(1 - a + b)z} + \mathcal{O}(\frac{1}{z^2}) \right] \label{eq:expansion}
\end{eqnarray}
for $|z| \to \infty$ and $a - b \notin \mathbb{Z}$, where $\Gamma(x) = \int_0^\infty t^{x - 1}\mathrm{e}^{-t}dt$ is the gamma function. Replacing $a = 1/2$, $b = (\alpha + 1)/2$, $c = (\alpha + 3)/2$ and $z = -1/(\lambda r)^2$ and doing some algebra, we finally find
\begin{equation}
\mathrm{Re} \{ I(r, \psi) \} \simeq \frac{\alpha + 1}{\alpha}\lambda r + \frac{\alpha + 1}{2(2 - \alpha)}(\lambda r)^3 + \frac{\Gamma \left( -\frac{\alpha}{2} \right) \Gamma \left( \frac{\alpha + 3}{2} \right)}{\sqrt{\pi}}(\lambda r)^{\alpha + 1}.
\label{eq:self_expanded}
\end{equation}
By imposing $r = \mathrm{Re} \{ I(r, \psi) \}$, we obtain
\begin{equation}
\frac{\Gamma \left( -\frac{\alpha}{2} \right) \Gamma \left( \frac{\alpha + 3}{2} \right)}{\sqrt{\pi}}\lambda^{\alpha + 1}r^\alpha + \frac{\alpha + 1}{2(2 - \alpha)}\lambda^3r^2 + \frac{\lambda}{\lambda_c} = 1.
\label{eq:expansion_final}
\end{equation}
We can distinguish then two different regimes:
\begin{enumerate}
\item If $\alpha < 2$, the second term of the left-hand side of Eq.~\eqref{eq:expansion_final} can be neglected, giving rise to
\begin{equation}
r(\lambda) \simeq \left[ \frac{\sqrt{\pi}}{\Gamma \left( -\frac{\alpha}{2} \right)\Gamma \left( \frac{\alpha + 3}{2} \right)} \frac{1}{\lambda^{\alpha + 1}} \left( 1 - \frac{\lambda}{\lambda_c} \right) \right]^\frac{1}{\alpha} = \frac{1}{\lambda} \left[ \frac{(\alpha + 1)\sqrt{\pi}}{\alpha \left|\Gamma \left( -\frac{\alpha}{2} \right) \right| \Gamma \left( \frac{\alpha + 3}{2} \right) \lambda} (\lambda - \lambda_c) \right]^\frac{1}{\alpha} \sim (\lambda - \lambda_c)^{1/\alpha}.
\label{eq:regime_1}
\end{equation}
\item If $\alpha > 2$, the first term of the left-hand side of Eq.~\eqref{eq:expansion_final} can be neglected, giving rise to
\begin{equation}
r(\lambda) \simeq \left[ \frac{2(2 - \alpha)}{\alpha + 1} \frac{1}{\lambda^3} \left( 1 - \frac{\lambda}{\lambda_c} \right) \right]^\frac{1}{2} = \frac{1}{\lambda} \left[ \frac{2(\alpha - 2)}{\alpha \lambda} (\lambda - \lambda_c) \right]^\frac{1}{2} \sim (\lambda - \lambda_c)^{1/2}. \label{eq:regime_2}
\end{equation}
\end{enumerate}

\subsubsection{Special cases}
\label{sec:special_cases}

Furthermore, we find analytical solutions of $r(\lambda)$ for two special cases:
\begin{enumerate}
\item $\alpha = 1$: The conviction distribution reads as $P(\rho) = 2\rho$, and the real part of the self-consistent function $I(r, \psi)$ is then
\begin{equation}
\mathrm{Re} \{I(r, \psi) \} = 2 \lambda r \int_0^1 d\rho \frac{\rho}{\sqrt{(\lambda r)^2 + \rho^2}} = 2 \lambda r \left[ \sqrt{(\lambda r)^2 + 1} - \lambda r \right].
\label{eq:self_1}
\end{equation}
Solving self-consistently $r = \mathrm{Re} \{I(r, \psi) \}$ we finally find
\begin{equation}
r(\lambda) = 1 - \left( \frac{1}{2\lambda} \right)^2,
\label{eq:solution_1}
\end{equation}
which, following the typical behavior $r(\lambda) \sim \left( \lambda - \lambda_c \right)^\beta$, corresponds to $\beta = 1$ with $\lambda_c = 1/2$.
\item $\alpha = \infty$: The conviction distribution can be written as $P(\rho) = \delta(\rho - 1)$. The real part of the self-consistent function is then
\begin{equation}
\mathrm{Re} \{ I(r, \psi) \} = \lambda r \int_0^1 d\rho \frac{\delta(\rho - 1)}{\sqrt{(\lambda r)^2 + \rho^2}} = \frac{\lambda r}{\sqrt{(\lambda r)^2 + 1}},
\label{eq:self_infty}
\end{equation}
which leads to
\begin{equation}
r(\lambda) = \sqrt{1 - \frac{1}{\lambda^2}},
\label{eq:solution_infty}
\end{equation}
corresponding to $\beta = 1/2$ with $\lambda_c = 1$.
\end{enumerate}
Interestingly, both cases show the same analytical solution of the order parameter
\begin{equation}
r(\lambda) = \left[ 1 - \left( \frac{\lambda_c}{\lambda} \right)^2 \right]^\beta
\label{eq:analytic_solution}
\end{equation}
different from zero for the depolarized state ($\lambda > \lambda_c$). In fact, at $\lambda = \lambda_c$ the new non-zero branch given by Eq.~\eqref{eq:analytic_solution} bifurcates from the trivial solution $r = 0$, expecting the usual transfer of stability of a supercritical pitchfork bifurcation.

\section{Solution for uncorrelated polarization: quadrimodal distribution}
\label{sec:uncorrelated_quadrimodal}

\begin{figure}[b!]
     \centering
\includegraphics[width=0.35\textwidth]{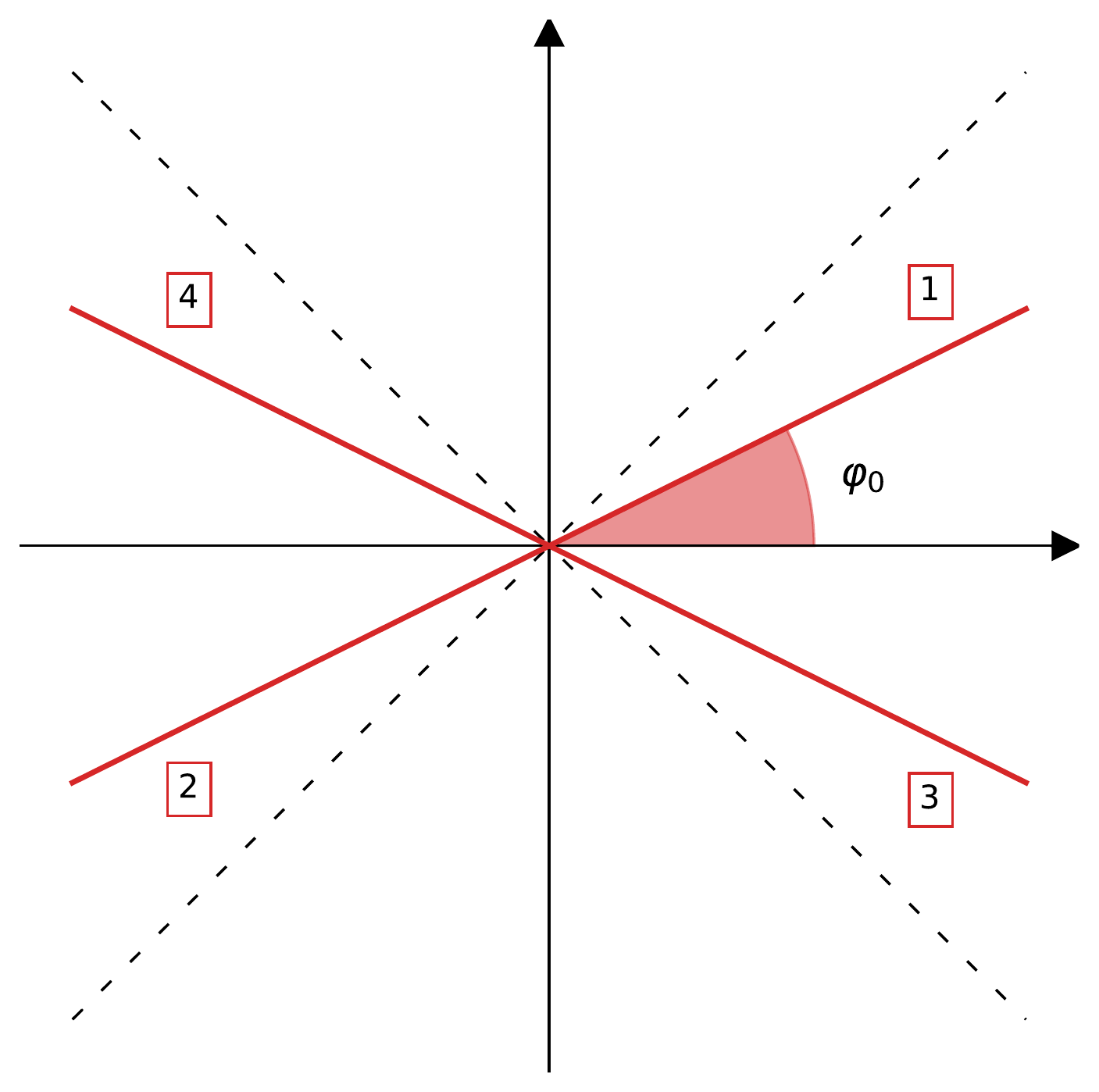} \hspace{3.0mm}
\includegraphics[width=0.35\textwidth]{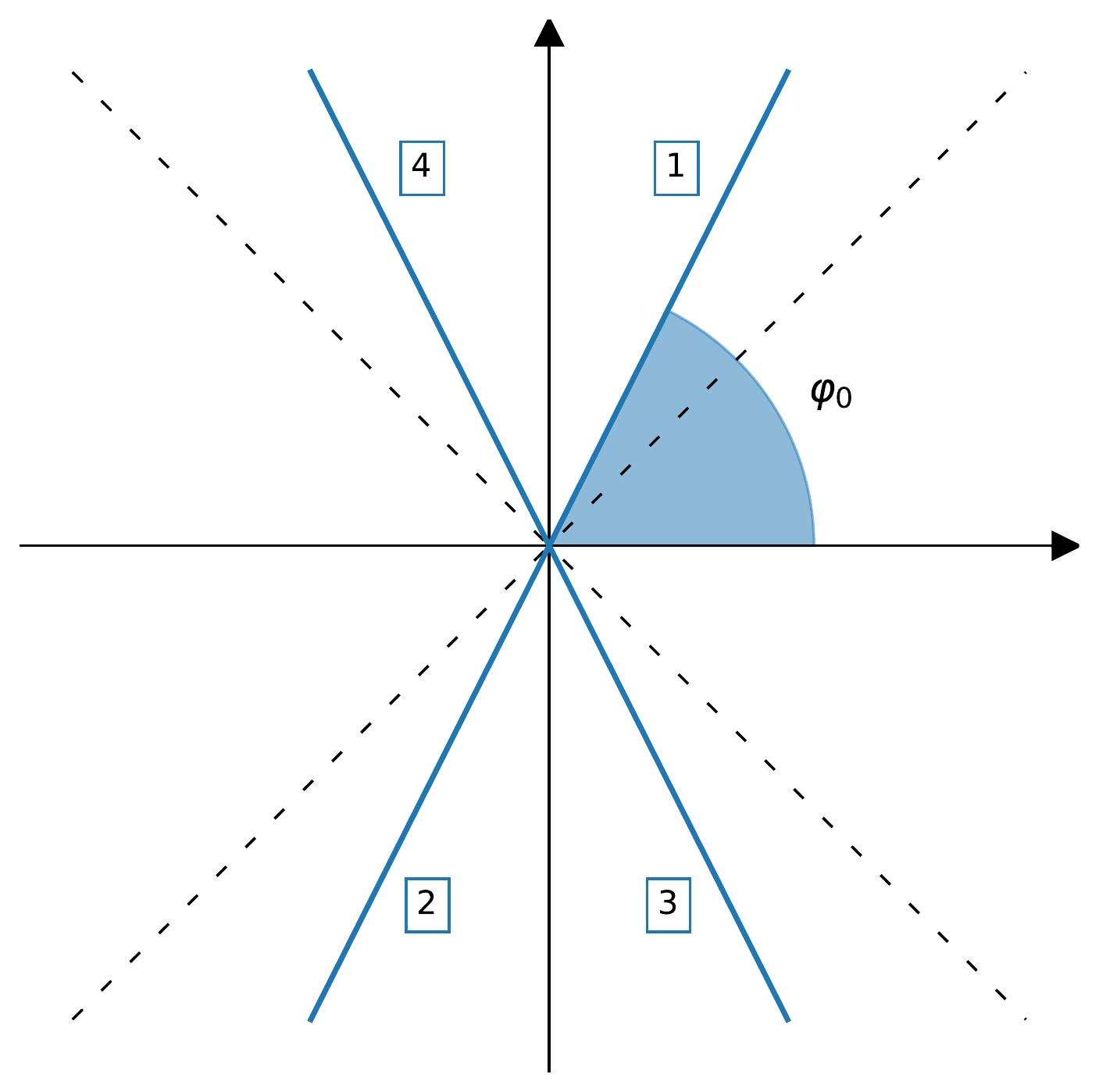}
     
        \caption{Schematic disposition of a quadrimodal distribution in the $xy$-plane for $\varphi_0 < \pi/4$ (left) and $\varphi_0 > \pi/4$ (right), where the case $\varphi_0 = \pi/4$ is shown in dashed lines. The four peaks are labelled in order of appearance in Eqs.~\eqref{eq:quadrimodal} and \eqref{eq:quadrimodal_integral}.}
        \label{fig:quadri_polar}
\end{figure}

A quadrimodal distribution counts on four peaks oriented in a particular way in the polar plane. It can be defined as
\begin{equation}
P(\varphi) = \frac{1}{4} \left[ \delta(\varphi - \varphi_0) + \delta(\varphi - \varphi_0 + \pi) \right] + \frac{1}{4} \left[ \delta(\varphi + \varphi_0) + \delta(\varphi + \varphi_0 - \pi) \right], \label{eq:quadrimodal}
\end{equation}
which corresponds to Eq.~(6) of the main text.
Notice, however, that here we explore the more general case with $\varphi_0 \in (0, \pi/2)$, which is necessary to numerically evaluate the value of $\psi$, as we will show later.
In \ref{fig:quadri_polar} we help the reader to visualize the $P(\varphi)$ for $\varphi_0 < \pi/4$ and $\varphi_0 > \pi/4$.

Ignoring the irrelevant normalization factor $1/4$ of the $P(\varphi)$ distribution, its integration in Eq.~\eqref{eq:angular_integral} gives rise to the four terms

\begin{eqnarray}
\Theta (\rho) &=& \frac{\sin(\varphi_0 - \psi)}{\sqrt{(\lambda r)^2 + 2\lambda r \rho \cos(\varphi_0 - \psi) + \rho^2}} - \frac{\sin(\varphi_0 - \psi)}{\sqrt{(\lambda r)^2 - 2\lambda r \rho \cos(\varphi_0 - \psi) + \rho^2}} \nonumber \\
&-& \frac{\sin(\varphi_0 + \psi)}{\sqrt{(\lambda r)^2 + 2\lambda r \rho \cos(\varphi_0 + \psi) + \rho^2}} + \frac{\sin(\varphi_0 + \psi)}{\sqrt{(\lambda r)^2 - 2\lambda r \rho \cos(\varphi_0 + \psi) + \rho^2}} \label{eq:quadrimodal_integral},
\end{eqnarray}
which are more difficult to analyze analytically than the bimodal case.
However, we can study the behavior of $\Theta (\rho)$ numerically for different values of $\lambda r$, $\varphi_0$ and $\psi$.
To do so, we have to realize that, in a similar way to the case of correlated polarization, $\psi$ can not take any value to preserve the polarization, i.e., ensure $r = 0$ as a solution.
From \ref{fig:quadri_polar}, we can distinguish two cases: $\varphi_0 \le \pi/4$ and $\varphi_0 \ge \pi/4$.
In the first case, we expect $\varphi_0 < \psi < \pi - \varphi_0$ (between the first and fourth peaks, respectively) or $\varphi_0 - \pi < \psi < -\varphi_0$ (second and third peaks).
In the second case, we expect $-\varphi_0 < \psi < \varphi_0$ (third and first peaks) or $\pi - \varphi_0 < \psi < \varphi_0 - \pi$ (fourth and second peaks).

\begin{figure}[tbp]
    \centering
    \includegraphics[width=0.9\textwidth]{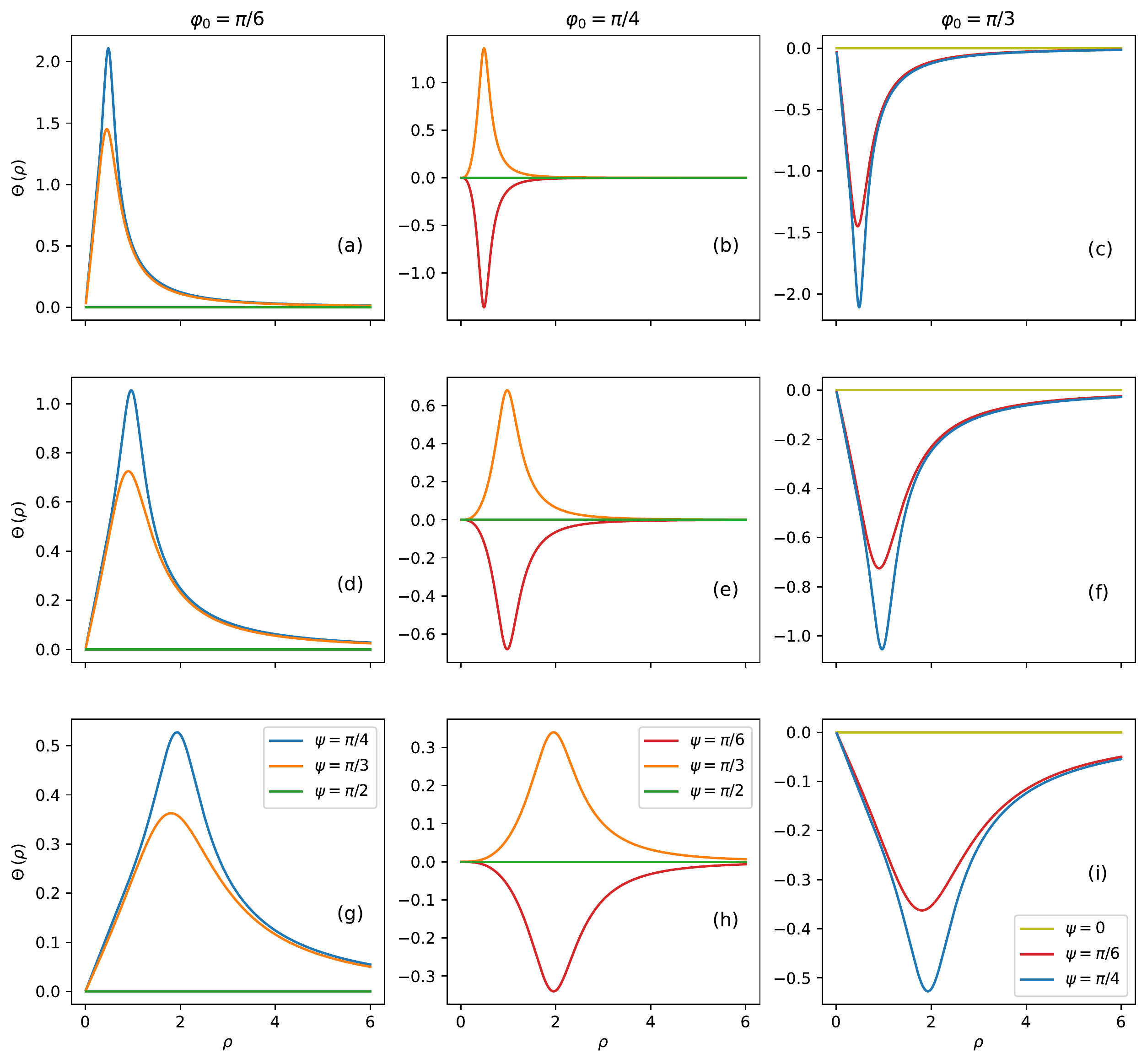}
    \caption{Numerical $\Theta$ as a function of $\rho$ for an uncorrelated polarization and different values of $\varphi_0$ and $\psi$. Different values of $\lambda r$ are also used: [(a), (b), (c)] $\lambda r = 0.5$, [(d), (e), (f)] $\lambda r = 1.0$, [(g), (h), (i)] $\lambda r = 2.0$.}
    \label{fig:quadri_numerical}
\end{figure}

Hence, in \ref{fig:quadri_numerical} we show the numerical results of $\Theta (\rho)$ for different values of the rest of the parameters. Indeed, it never changes sign, being always positive, negative or zero for any $\rho, \lambda \in [0, \infty]$.
Now we can infer the value of $\psi$ by means of $\Theta = 0$. This is satisfied if and only if the four terms of Eq.~\eqref{eq:quadrimodal_integral} cancel in pairs. There are another time two possibilities:
\begin{enumerate}
\item The first and second terms are cancelled by the fourth and third ones, respectively. This is the case of $\varphi_0 \le \pi/4$, for which the peak cancellation occurs along the $x$-axis. Thus, $\sin(\varphi_0 - \psi) = - \sin(\varphi_0 + \psi)$ and $\cos(\varphi_0 - \psi) = - \cos(\varphi_0 + \psi)$.
\item The first and second terms are cancelled by the third and fourth ones, respectively. This is the case of $\varphi_0 \ge \pi/4$, for which the peak cancellation occurs along the $y$-axis. Thus, $\sin(\varphi_0 - \psi) = \sin(\varphi_0 + \psi)$ and $\cos(\varphi_0 - \psi) = \cos(\varphi_0 + \psi)$.
\end{enumerate}
To summarize it, we obtain
\begin{equation}
\psi =
    \begin{cases}
        \pm \pi/2 & \text{if } \varphi_0 \le \pi/4 \\
        0, \pm \pi & \text{if } \varphi_0 \ge \pi/4
    \end{cases},
\label{eq:quadrimodal_psi}
\end{equation}
in full agreement with the results of \ref{fig:quadri_numerical}. Since the definition given by Eq.~(6) of the main text is restricted to $\varphi_0 \in (0, \pi/4]$ due to the angular symmetry, the result is reduced to $\psi = \pm \pi/2$. 
This effectively coincides with what was obtained also for a correlated polarization, so in general for $\varphi_0 \in [0, \pi/4]$ the average orientation is $\psi = \pm \pi/2$.

\subsection{Power-law distributed conviction}
\label{sec:uncorrelated_power}

\begin{figure}[tbp]
    \centering
    \includegraphics[width=0.5\textwidth]{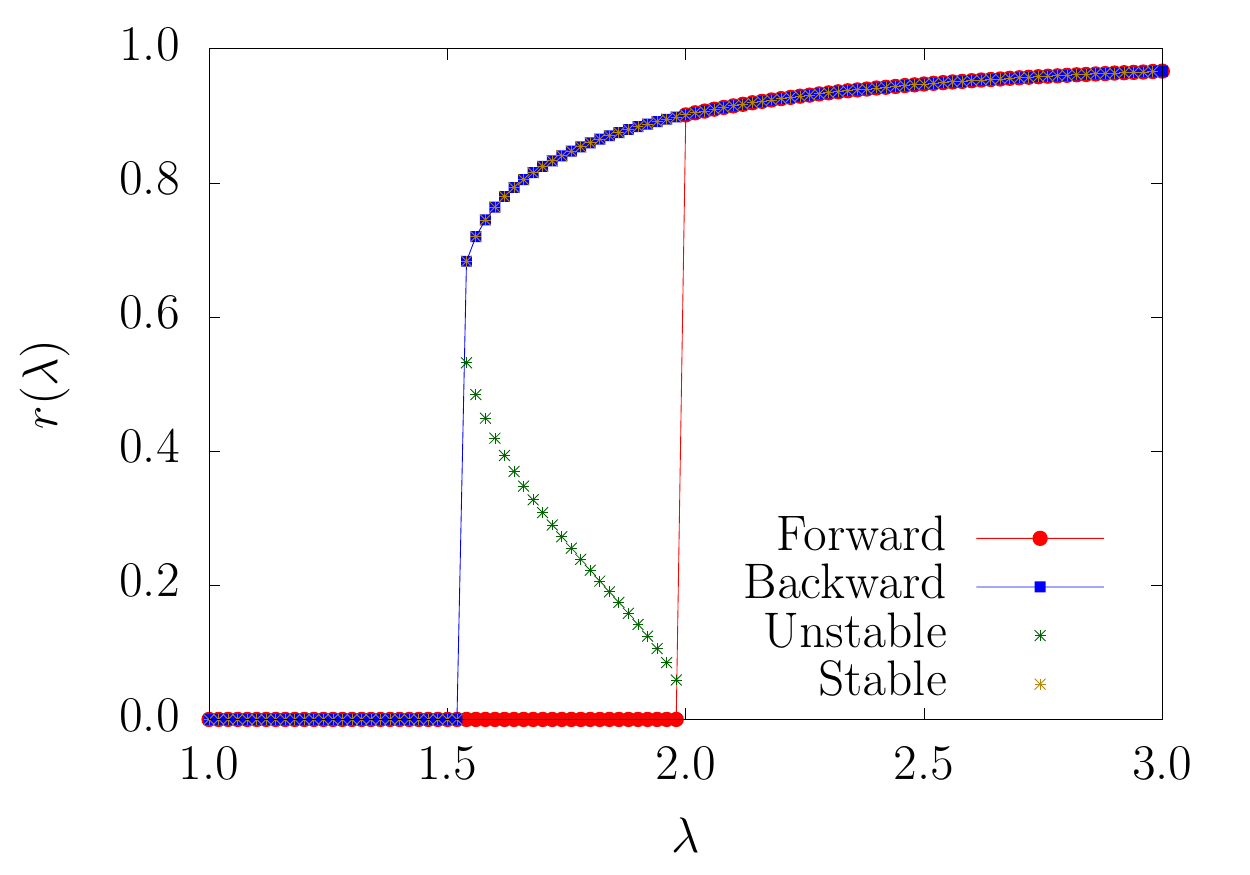}
    \caption{Order parameter $r$ as a function of $\lambda$ in a system of size $N = 10^3$ for an uncorrelated polarization with $\varphi_0 = \pi/4$ and a conviction distribution with $\alpha = \infty$. Forward and backward continuations in $\lambda$ are computed, as well as the theoretical prediction through star points for the stable and unstable solutions.}
    \label{fig:hysteresis}
\end{figure}

The general definition of the distribution given by Eq.~\eqref{eq:quadrimodal} (Eq.~(6) of the main text) does not allow us to determine analytically the modulus $r$. In particular, the real part of the self-consistent function $I(r, \psi)$ of a perpendicular quadrimodal distribution with four peaks separated each other by the same angle $\varphi_0 = \pi/4$ can be easily obtained from Eq.~\eqref{eq:real_part}, taking the form
\begin{equation}
\mathrm{Re} \{ I(r, \psi) \} = \frac{\alpha + 1}{2} \int_0^1 d\rho \frac{\left( \lambda r + \frac{\sqrt{2}}{2} \right) \rho^\alpha}{\sqrt{(\lambda r)^2 + \sqrt{2} \lambda r \rho + \rho^2}} + \frac{\alpha + 1}{2} \int_0^1 d\rho \frac{\left( \lambda r - \frac{\sqrt{2}}{2} \right) \rho^\alpha}{\sqrt{(\lambda r)^2 - \sqrt{2} \lambda r \rho + \rho^2}}.
\label{eq:self_quadrimodal}
\end{equation}
Its numerical integration and subsequent solution of $r = \mathrm{Re} \{ I(r, \psi) \}$ allow us to predict numerically the nature of the phase transition. Interestingly, as $\alpha \to \infty$ the conviction distribution approaches $P(\rho) = \delta(\rho - 1)$, giving rise to the hysteresis phenomenon shown in \ref{fig:hysteresis}. The transition turns out to be of first-order, characterized by the presence of metastability in the solution of $r$.

\subsection{Critical angle}
\label{sec:uncorrelated_critical}

The case of $\alpha = \infty$ is especially interesting by the fact that, for $\varphi_0 = \pi/4$, the hysteresis effect is really pronounced, being the difference between thresholds following a forward and backward continuations in $\lambda$ maximum regarding the rest of the values of $\alpha$. For this reason, determining the critical angle for which the phase transition goes from being continuous to explosive becomes important. Analogously to Eq.~\eqref{eq:self_quadrimodal}, the real part of the self-consistent function $I(r, \psi)$ in terms of $\varphi_0$ follows as
\begin{equation}
\mathrm{Re} \{ I(r, \psi) \} = \frac{\lambda r + \sin{(\varphi_0)}}{2 \sqrt{(\lambda r)^2 + 2 \lambda r \sin{(\varphi_0)} + 1}} + \frac{\lambda r - \sin{(\varphi_0)}}{2 \sqrt{(\lambda r)^2 - 2 \lambda r \sin{(\varphi_0)} + 1}}.
\label{eq:self_angle}
\end{equation}
Expanding $\mathrm{Re} \{ I(r, \psi) \}$ at $r \to 0$, we can investigate which is the non-zero solution that is found above the critical point. Obviously, $\mathrm{Re} \{ I(0, \psi) \} = 0$ according to the definition of polarized state. Moreover, at $r = 0$ we find an inflection point, i.e., $\partial ^2 \mathrm{Re} \{ I(r, \psi) \} / \partial r^2 \vert_{r=0} = 0$. Thus, we write
\begin{equation}
\mathrm{Re} \{ I(r, \psi) \} \simeq r \left. \frac{\partial \mathrm{Re} \{ I(r, \psi) \}}{\partial r} \right|_{r = 0} + \frac{r^3}{6} \left. \frac{\partial ^3 \mathrm{Re} \{ I(r, \psi) \}}{\partial r^3} \right|_{r = 0}.
\label{eq:angle_expanded}
\end{equation}
Doing the algebra and ignoring the zero solution once imposed $r = \mathrm{Re} \{ I(r, \psi) \}$, this leads to
\begin{equation}
\frac{\lambda^3}{2} \left[ 6\sin^2{(\varphi_0)} - 5\sin^4{(\varphi_0)} - 1 \right] r^2 + \frac{\lambda}{\lambda_c} = 1.
\label{eq:expansion_angle}
\end{equation}
Since $\varphi_0 \in [0, \pi/4]$, the first term of the left-hand side of the previous equation vanishes at $\varphi_c = \arcsin{\left( 1/\sqrt{5} \right)}$. Therefore, a sign change is found as
\begin{equation}
r(\lambda) \sim
    \begin{cases}
        (\lambda - \lambda_c)^{1/2} & \text{if } \varphi_0 < \varphi_c \\
        (\lambda_c - \lambda)^{1/2} & \text{if } \varphi_0 > \varphi_c
    \end{cases},
\label{eq:angle_behavior}
\end{equation}
with a continuous transition (bifurcation branch) for the first case and an explosive one (hysteresis unstable solution, see \ref{fig:hysteresis}) for the second case.

\section{Simulation details}

In order to check for the presence of hysteresis in explosive transition, we perform a set of forward and backward continuation simulations. In the forward continuation scheme, we perform simulations starting from a very small value of $\lambda$, and run them until the steady state is reached. The value of lambda is then slightly increased and the simulation is run again. In this scheme we are exploring the lower branch of the solution, and therefore the value of $\lambda$ at which we start observing a non-zero solution corresponds to the threshold $\lambda_c$. In the backward continuation scheme we start instead with a large value of $\lambda$ and subsequently reduce it. In this kind of experiment we are exploring the upper branch of the transition and the point at which a zero solution is first observed corresponds to the threshold $\lambda_u$ of the upper branch. For continuous, second order transitions we expect $\lambda_u = \lambda_c$, while for an explosive, discontinuous transition we have $\lambda_u < \lambda_c$.

\section{Empirical distributions}
\label{sec:empirical}

\begin{figure}[tbp]
    \centering
    \includegraphics[width=0.9\textwidth]{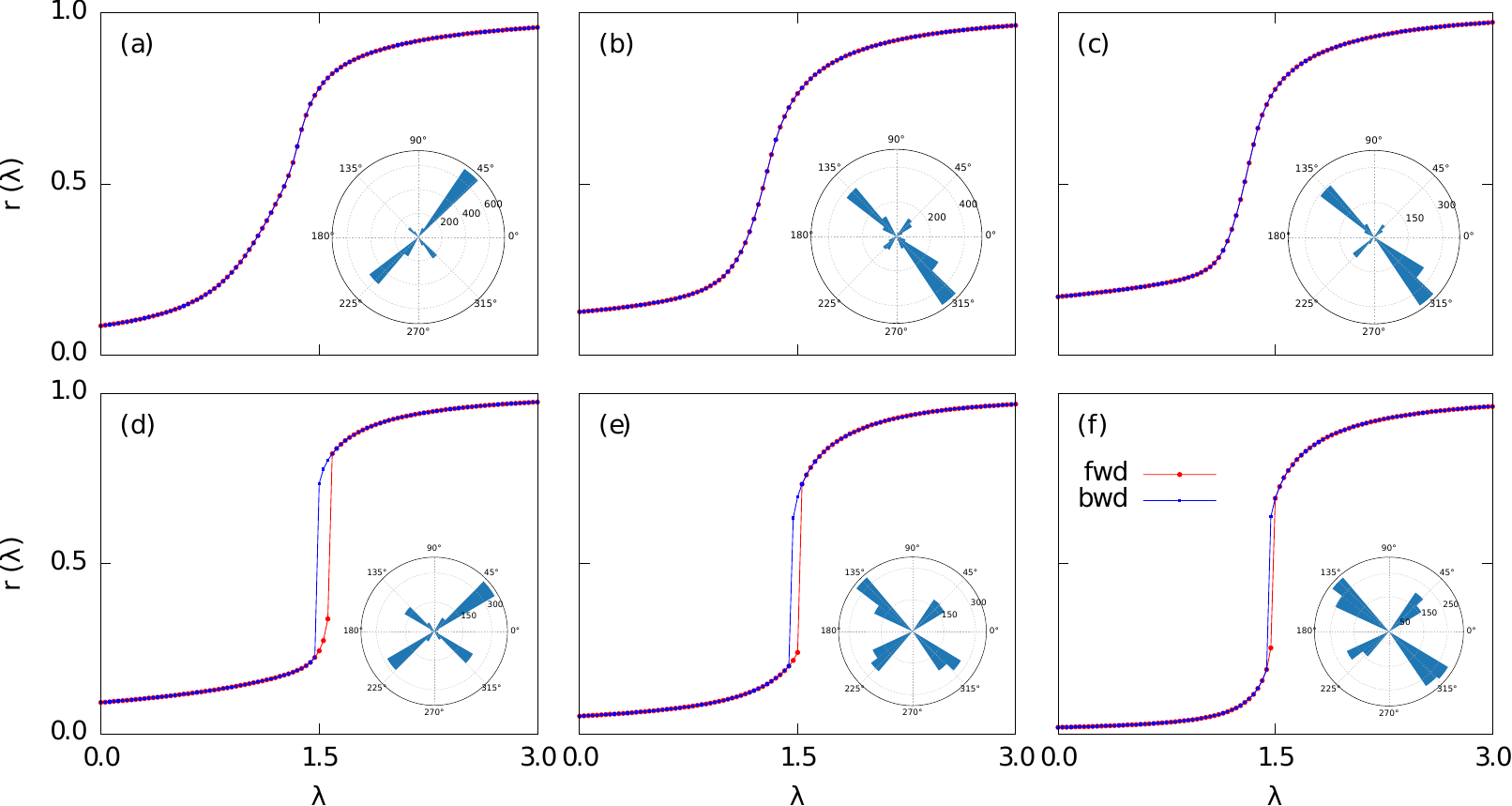}
    \caption{Main: Order parameter $r$ as a function of $\lambda$ including both forward (fwd) and backward (bwd) continuations in $\lambda$. Inset: Real ANES opinions about two distinct topics represented in polar coordinates at $\lambda = 0$. These empirical distributions are obtained neglecting every agent with conviction lower than the median of $P(\rho)$, approaching then the case $\alpha \to \infty$. Simulations are performed for empirical correlated [(a) \texttt{Mexican wall} and \texttt{same sex}, (b) \texttt{obamacare} and \texttt{same sex}, (c) \texttt{obamacare} and \texttt{transgender}] and uncorrelated [(d) \texttt{obamacare} and \texttt{religion}, (e) \texttt{climate change} and \texttt{fight ISIS}, (f) \texttt{religion} and \texttt{fight ISIS}] opinion distributions.}
    \label{fig:order_transition}
\end{figure}

In relation to Fig.~4 of the main text, in \ref{fig:order_transition} we provide more examples of first- and second-order transitions (bottom and top row, respectively) for the case $\alpha = \infty$ depending on whether the empirical $P(\varphi)$ is uncorrelated or correlated.

\end{document}